\documentclass[prb,reprint]{revtex4-2}

\usepackage{amsmath, amssymb}
\usepackage[normalem]{ulem}

\usepackage{graphicx}
\graphicspath{{./figures/}{../process/}}

\usepackage[T1]{fontenc}
\usepackage[utf8x]{inputenc}
\usepackage{lmodern}

\usepackage{xcolor}
\usepackage{multirow}  
\usepackage{textcomp}  

\usepackage{siunitx}

\usepackage{hyperref}

\newcommand{\bnabla}{{\mbox{\boldmath $\nabla$}}}
\newcommand{\bphi}{{\mbox{\boldmath $\varphi$}}}
\newcommand{\vvn}{{\bf {v}}_n}
\newcommand{\vvs}{{\bf {v}}_s}

\begin{document}

\title{Cooling with a subsonic flow of quantum fluid}

\author{Pantxo Diribarne}
\email{pantxo.diribarne@univ-grenoble-alpes.fr}
\author{Bernard Rousset}
\affiliation{Univ. Grenoble Alpes, CEA IRIG-DSBT, 38000 Grenoble, France}
\author{Yuri A. Sergeev}
\affiliation{Joint Quantum Centre Durham-Newcastle, School of
   Mathematics, Statistics and Physics, Newcastle University, Newcastle
   upon Tyne, NE1 7RU, UK}
\author{Camille Noûs}
\affiliation{Laboratoire Cogitamus, France}
\author{Jérôme Valentin}
\altaffiliation[present address ]{LERMA, Observatoire de Paris, 75014 Paris}
\author{Philippe-Emmanuel Roche}
\affiliation{Univ. Grenoble Alpes, CNRS, Institut NEEL, F-38042 Grenoble, France}


\begin{abstract}
  Miniature heaters are immersed in flows of quantum fluid and
  the efficiency of heat transfer is monitored versus velocity,
  superfluid fraction and time. The fluid is $^4$He helium with a
  superfluid fraction varied from 71\% down to 0\% and an imposed velocity up to
  \SI{3}{m/s}, while the characteristic sizes of heaters range from
  $\SI{1.3}{\mu m}$ up to a few hundreds of microns.
  At low heat fluxes, no velocity dependence is observed, in agreement
  with expectations. In contrast, some velocity dependence emerges at
  larger heat flux, as reported previously, and three
  nontrivial properties of heat transfer are identified. 

  First, at the largest  superfluid fraction (71\%), a new heat
  transfer regime appears at non-null velocities and it is typically
  $10\%$ \textit{less} conductive than at zero velocity.
  Second, the velocity dependence of the mean heat transfer is compatible
  with the square-root dependence observed in classical
  fluids. Surprisingly, the prefactor to this dependence is maximum
  for an intermediate superfluid fraction or temperature (around
  \SI{2}{K}). Third, the heat transfer 
  time series exhibit highly conductive short-lived events. These
  \textit{cooling glitches} have a velocity-dependent characteristic
  time, which manifest itself as a broad and energetic peak in the
  spectrum of heat transfer time series, in the kHz range.
 
  After showing that the velocity dependence can be attributed to the 
  breaking of superfluidity within a thin shell surrounding heaters,
  an analytical model of  forced heat transfer in a quantum flow is
  developed to account for the properties reported above. We argue that
  large scale flow patterns must form around the heater, having a size
  proportional to the heat flux (here two decades larger than the
  heater diameter) and resulting in a turbulent wake. 
  The observed spectral peaking of heat transfer is quantitatively
  consistent with the formation of a Von Kármán vortex street in
  the wake of a bluff body nearly two decades larger than the heater
  but its precise temperature and velocity dependence remains unexplained.
  An alternative interpretation for the spectral peaking is discussed,
  in connection with existing predictions of a bottleneck in the
  superfluid velocity spectra and energy equipartition.

\end{abstract}

\maketitle

\section{Introduction and motivation}

Below its superfluid transition temperature, liquid helium $^4$He enters
the He II phase which displays amazing quantum properties at large
scales~\cite{donnellyPhysToday2009}. In particular, this fluid can flow
without viscous friction, it hosts propagating heat waves 
-- called second sound waves -- and is extremely efficient in transporting heat.

Quantum fluids~\cite{barenghi2016primer} such as He II are also characterized
by the existence of quantized vortex filaments which concentrate
all the vorticity of the superfluid~\cite{DonnellyLivreVortices}.
Although the presence of superfluid vortices reduces the efficiency
of heat transport, the latter remains much more efficient than standard
convection and diffusion heat transport in most situations~\cite{VanSciverLivre2012}.

A famous model to describe heat transport and hydrodynamics of quantum
fluids at finite temperature is Tisza and Landau's two-fluid model,
which describes He~II as an intimated mixture of an inviscid 
\textit{superfluid component} and a viscous \textit{normal component}
that contains all the entropy of the fluid~\cite{Khalatnikov65}. 
The local relative density fraction of both components depends on the local
temperature. 

Thus, steady heat transport can be described as a flow of normal component
carrying its entropy. When this mass flow is balanced by an opposite mass
flow of superfluid, we have a so-called \textit{thermal counter-flow}.
This situation occurs, for instance, in the vicinity of heaters and coolers,
and it has been extensively studied in pipe and channel
geometries~\cite{VanSciverLivre2012}.

Contrary to the situation in classical fluids, forced convection in He~II
has long been assumed not to improve measurably heat transfer,
because the classical convection and diffusion mechanisms are far
less efficient than counter-flows in transporting heat, at least in
subsonic flows. The special case of flows reaching or exceeding the velocity
of second sound, or even first sound in helium (typically \num{16.5}
and \SI{227}{m/s} at \SI{2}{K}) is not addressed in the present study,
nor in others to the best of our knowledge. 

Yet, in a recent instrumental study, \textcite{Duri15} reported that an external
(subsonic) flow can favor heat transfer from a hot-wire, but the underlying
mechanism was not addressed.

This paper reports en experimental study of forced heat transfer from heaters
immersed in a subsonic flow of superfluid, and reveals a rich phenomenology.

Related previous studies are reviewed in Sec.~\ref{sec:soa}. 
The experiments are presented in Sec.~\ref{sec:experiment},
in particular the subsonic flows and the various miniature heaters
used. Sections \ref{sec:bistable}, \ref{sec:velocityeffect} and
\ref{sec:coolingglitches} report three key properties of forced heat
transfer in He~II: 
the existence of metastable conduction states, velocity and temperature
dependencies of heat transfer, and the existence of short-lived
cooling events, named cooling glitches. Section.~\ref{sec:discussion}
presents analytical models accounting for some -- but not all -- observations.

\section{State of the art}
\label{sec:soa}
In the absence of an external flow, He~II heat transfer studies
are often reported in the thermal counter-flow literature. In
particular, the modeling of nonplanar geometries has recently been the
subject of a number of numerical and theoretical studies, most of
which predict non trivial behaviors.

\textcite{Saluto14} have used a so-called \textit{hydrodynamical model}~\cite{Mongiovi07}
to assess the behavior of the vortex line density of a counter-flow between
two concentric cylinders at different temperatures. From their initial
model they derived a modified Vinen equation which, in addition to the original
source and sink terms, features a vortex diffusion term. In the presence of a
nonuniform heat-flux, the model predicts a nonuniform vortex line density
(as does the original Vinen model) with a diffusive migration of
vortices produced  in the most dense region to the most dilute region.
The main consequence of this addition is that if the heat flux is varied faster
than the typical diffusion time, the local vortex line density has an
hysteretic behavior.

Using the vortex filament method \textcite{Varga19} has shown that in spherical
geometry (using a point source), for bath temperatures larger than \SI{1.5}{K},
all initial seeding vortices are annihilated on the virtual heat
source. For smaller  temperatures, a self-sustained vortex tangle was generated
but, due to  computational limitations, it could not reach a stationary state.
\textcite{Inui20} have run a similar numerical simulation with a different approach for
the core: instead of a point source, they simulated an actual spherical heater
(of a finite diameter) using suitable boundary conditions for the normal and superfluid
velocities. Contrary to \textcite{Varga19} they show that they
are able to obtain a self-sustained vortex tangle at most temperatures,
with a non trivial density profile.

\textcite{Rickinson20} used the same vortex filament method to model the vortex tangle of
a cylindrical counter-flow, with a finite inner diameter. What they find
is that in order to reach a stationary state, they need to specify
a radius dependent friction parameter between the two components of He~II
(which somewhat mimics the effect of an actual temperature gradient).
The latter trick was inspired by a previous finding \cite{Sergeev19},
that showed that using the coarse-grained Hall-Vinen-Bekarevich-Khalatnikov 
(HVBK) model it was necessary to
take the variations of the fluid properties around the wire into account,
in order to reach a stationary state in cylindrical geometry.
\textcite{Rickinson20} showed the standard scaling
for the vortex line density $\mathcal{L}$ as a function of the relative velocity
$v_{ns}$ between the two components holds: $\mathcal{L}\propto v_{ns}^n$ with
$n\approx 2$. This is an important result in that it allows for the use of standard
macroscopic laws for the heat transfer around non planar surfaces. Among others,
it supports \textit{a posteriori} the use of the conduction function
when simulating the heat flux around a cylindrical heater~\cite{Duri15}.

Now we turn to the problem of heat transfer in He~II in the presence of an
external flow, for which the literature is much sparser. First,
two experimental studies in pipe flows are worth mentioning.
\textcite{Johnson78} have measured the heat flux through a tube in the
presence of both temperature and pressure gradients
and concluded that the presence of a
pressure driven flow inside  the tube somewhat increased the mutual
friction between the superfluid  and normal components, thereby
depleting the efficiency of the heat transfer.
\textcite{Rousset92} measured the temperature profile around a heater
that was placed in the middle of a tube traversed by a subsonic He~II
flow. They were able to account for most of the results using simple
entropy conservation model and  isenthalpic expansion corrections
(see also Refs.~[\onlinecite{Rousset94,Fuzier2001,Fuzier:2008}]).

A third experimental observation is directly related to the present one.
In an instrumentation study, \textcite{Duri15} reported that an external flow
increases the heat transfer around a hot-wire anemometer, which is basically
an overheated wire-shaped thermometer.
They were able to account quantitatively for the heat transfer at null velocity
assuming that standard counter-flow laws still hold in cylindrical geometry despite
very high heat flux, but did not propose any explanation for the heat transfer
improvement due to the external flow.

To the best of our knowledge, there has not been any attempt at studying
specifically the effect of an external flow on the heat transfer at
the interface between a solid body and He~II. This paper attempts to fill these
gaps in our understanding of heat transfer in superfluid flows.

\section{Experimental conditions}
\label{sec:experiment}
This study uses three different miniature heaters immersed in flows of
He~II to assess the properties of intense heat transfer in subsonic
quantum flows. 

In the following we first describe the measurement protocol and then
provide the detailed description of the flows and, finally, of the heaters.

The experimental conditions are summarized in Table~\ref{tab:exp_cond}.

\begin{table}
  \begin{tabular}{|c|c|c|c|c|c|}
    \hline
    Heater&Facility / Flow&$P$ [\si{bar}]&$v_\infty$[m/s]&$T_\infty$ [K]&$X_{sf}$[\%]\\
    \hline
    Wire & HeJet /    &2.6& 0 - 0.40 & 1.74 & 71\\
         & grid flow  &   & 0 - 0.52 & 1.93 & 51\\
         &            &   & 0 - 0.52 & 2.05 & 31\\
         &            &   & 0 - 0.52 & 2.13 & 10\\
         &            &   & 0 - 0.52 & 2.29 & 0\\
    \hline
    Film & HeJet /    &2.6& 0.38     & 2.00 & 40\\
         &grid flow   &   &          &      & \\
    \hline
    Chip & SHREK /    &3.0& 0 - 3    & 1.6 - 2.1 & 82 - 20\\
     &  rotating flow &   &  (0 - 1.2~Hz)&  & \\
    \hline
  \end{tabular}
  \caption{Summary of experimental conditions for all heaters. Here $v_\infty$
    and $T_\infty$ are, respectively, the fluid velocity and temperature away 
    from the heater. The density fraction of superfluid component $X_{sf}$ is
    estimated in pressurized helium using the
    HEPAK\textsuperscript{\textregistered} library.}
  \label{tab:exp_cond}
\end{table}

\subsection{Measurement principles.}

The heat flux from the heaters is produced by the Joule effect, $\dot Q=eI
= RI^2$, where $e$ is the voltage across the heater, $I$ is the current through
 it, and $R$ is its electrical resistance. The spatially
averaged temperature of the heater $T_w$
is inferred from the calibration law $R(T_w)$ of its
temperature-dependent resistance. The heaters can thus be considered
as overheated thermometers. Their different shapes and sizes are
described in Sec.~\ref{sec:heaters}.

In order to monitor the fluctuations of the heat transfer, two types
of electronics circuitry are used to drive the heaters:
constant-current sources and  a constant-resistance (or temperature)
controller. The latter is a commercial hot-wire anemometry controller
able to control the resistance over a bandwidth exceeding DC-\SI{30}{kHz}
(DISA model 55-M10).  The measured voltage $e$ is either the
voltage drop across the heater when using the constant current
circuit, or an image of the current through the heater (via a shunt resistance)
when using the constant resistance controller. In both
cases, time series are calculated for the total heat flux $\dot Q$ and
the heater overheating  $T_w-T_\infty$ with respect to the fluid
temperature away from the heater, $T_\infty$.

The use of two types of electronics allows us to check if the observed
instabilities are artifacts associated with the electronic
circuitry. The signals are acquired by a delta-sigma analog-to-digital
converter (NI-PXI4462), at sampling frequencies up to 100 kHz (most
often 30~kHz). For given flow conditions, the typical data set
consists of 15 files with $4\times10^6$ data samples.

\subsection{Descriptions of the flows}
\label{sec:flows}
Two facilities in Grenoble, SHREK and HeJet, are used to produce
pressurized flows with a steady mean velocity and limited turbulent
fluctuations. The pressurization of the flow above the fluid critical
pressure is required to prevent boiling or the
formation of a gas film around the heater irrespective of the amount of overheating.

In mechanically-driven isothermal turbulent flows, such as those
produced by both facilities, the superfluid and the normal fluid
components that make up He~II  are locked together at large and
intermediate flow scales~\cite{Maurer98,Salort10}.
In the quantum turbulence literature, such flows are sometimes
referred to as co-flows, to distinguish them from the thermally driven
He~II flows, called counter-flows. Surely, as discussed later, the
flow in the close vicinity of the heater is no longer a co-flow.

In the following subsections we give the most important details about the HeJet
facility where most measurements were done using the two smallest
heaters, and the SHREK
facility, in which measurements with the largest heater were performed.

\subsubsection{HeJet: The grid flow}

\begin{figure}[t]
  \centering
  \includegraphics[]{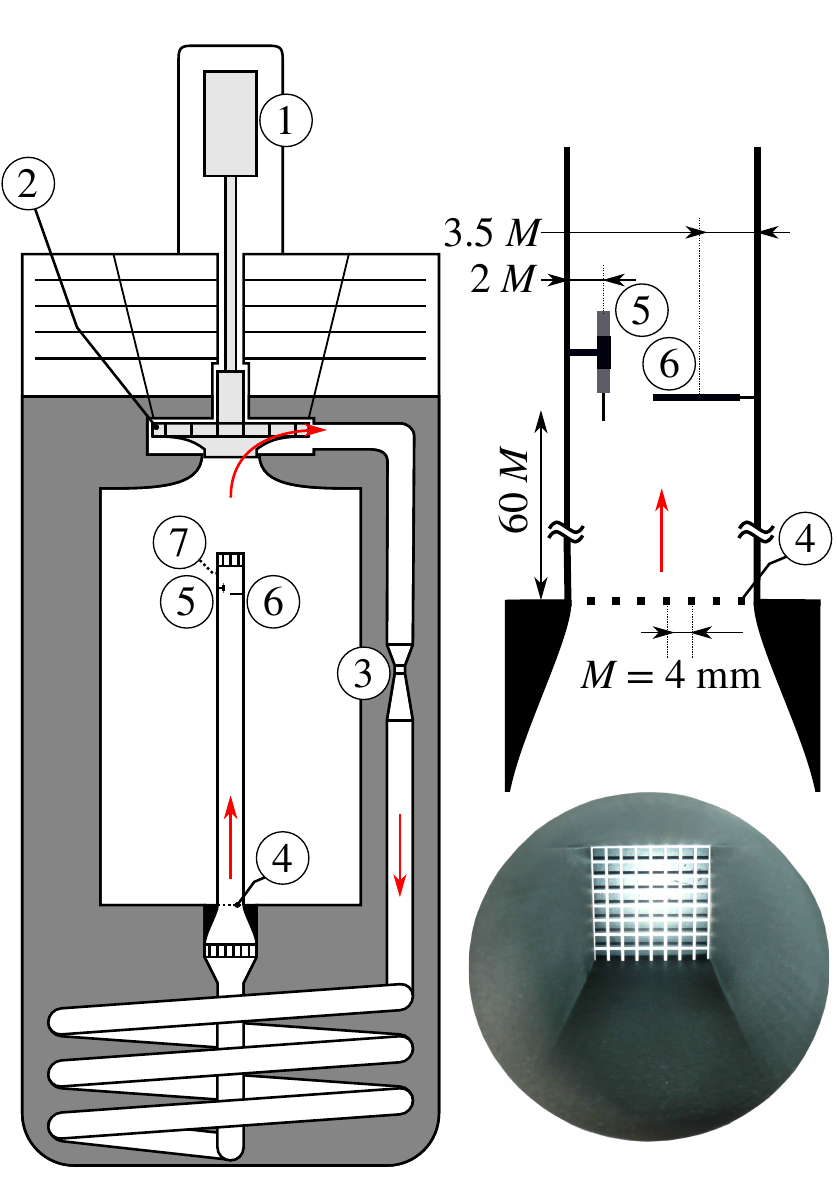}
  \caption{Left: Sketch of the experimental apparatus. \textbf{1}: DC motor.
    \textbf{2}: Centrifugal pump. \textbf{3}: Venturi
    flow-meter. \textbf{4}: Grid. \textbf{5}: Pt-Rh
    wire heater. \textbf{6}: Film heater array. \textbf{7}: Temperature sensor.
    Right: zoom of the test section with relevant dimensions and a
    picture of the convergent followed by the grid.}
  \label{fig:sketch}
\end{figure}

The HeJet facility is a closed loop of pressurized liquid helium
immersed in a liquid helium bath at saturated pressure (dark gray in
Fig.~\ref{fig:sketch}). The flow in the loop is driven by a
centrifugal pump empowered by a DC motor at room temperature. The
facility, originally designed to produce an inertial round jet of
liquid helium~\cite{Duri11}, has been modified to produce a
turbulent  grid flow (see Fig.~\ref{fig:sketch}). The motivation for
this change was to obtain a quantum flow with relative velocity fluctuations $I$
within a few percents.

The experimental flow section consists of a 32$\times$\SI{32}{mm^2} square
cross section  tunnel with length \SI{450}{mm}. Prior to entering the
tunnel, the flow goes through the conditioning section: a divergent (\SI{32}{mm}
to \SI{50}{mm} round section) followed by a 16~mm long honeycomb with
\SI{3}{mm} mesh size  and then a convergent part which smoothly
concentrates the flow into the  square tunnel section.

The grid is etched by wire electro-erosion in a \SI{0.8}{mm} thick
stainless steel plate. The rods are thus $\SI{0.8}{mm}\times
\SI{0.8}{mm}$ wide and the mesh size  is $M=\SI{4}{mm}$ which leads to
a solidity (or obstruction ratio) of 36\%. The geometry of the grid
follows the now standard recommendations from \textcite{Comte66}.

Measurements are done at a distance of 60~$M$ downstream the
grid. At this location, the longitudinal integral length scale is
$L_f = 5.0\pm \SI{0.2}{mm}$ and the turbulence intensity, 
defined as the ratio of the root-mean squared fluctuating velocity $v'$ to the
mean velocity $v_\infty$,is $I\approx 2.6\%$. 
The procedure for characterizing the flow is detailed in the Appendix.

The range of explored temperatures is $\SI{1.74}{K}$ to
$\SI{2.28}{K}$, corresponding to a superfluid fraction from 71\% to
0\%. The temperature in the pressurized bath is measured at the outlet
of the grid flow tube (see Fig.~\ref{fig:sketch}) with a Cernox\textsuperscript{\textregistered}
thermometer  and is regulated by means of a heater within a few tenths of milliKelvin.
The absolute value of the temperature, known to better than \SI{1}{mK},
is checked \textit{in situ} using the saturated pressure of the (superfluid) outer bath
when the pressurized flow is at rest.

The range of mean velocities is $v_\infty=0$ to
\SI{0.52}{m/s}, as calculated from the Venturi flow-meter pressure drops 
(see item 3 in Fig.~\ref{fig:sketch}).

For all experiments, the pressure is maintained at $2.6\pm\SI{0.1}{bars}$.
In such condition, the superfluid transition occurs at $T_\lambda
\approx \SI{2.15}{K}$.

\subsubsection{SHREK: The rotating flow}

SHREK is a large cylindrical vessel, $D_s = \SI{78}{cm}$
in inner diameter and \SI{116}{cm} in height, equipped with two
identical turbines facing each other (see \textcite{Rousset14} for
details). The turbines are fitted with curved blades so that, depending on
their respective rotation direction, the facility can produce different kinds
of flows: from the quasi solid rotation flow when turbines rotate in
the same direction (co-rotation), to the von K\'arm\'an  flow when
turbines rotate in opposite directions (counter-rotation).

In this paper we report data acquired in co-rotation from a bare
chip heater  (see Sec.~\ref{sec:chip}) located in the mid
plane of the vessel, 1~cm away from the wall. This sensor was
previously used as an anemometer in He~I (see Fig.~14 in
Ref.~[\onlinecite{Rousset14}]). In those co-rotation conditions, the turbulence intensity
was found to be of the order $5\%$.

In order to estimate the velocity of the fluid around the sensor, we
assume that the co-rotation produces a solid-body rotation
flow with the same angular velocity $\omega$ as the turbines:
$v_\infty = \omega D_s/2 $. This simple model probably slightly overestimates
the  velocity but it gives an order of magnitude of the velocity with
sufficient accuracy for the purpose of the current study.

The flow pressure is maintained at $\SI{3}{bars}$ to avoid boiling and
cavitation on the miniature heaters. In such conditions, the
superfluid transition occurs also at $T_\lambda \approx \SI{2.15}{K}$.

\subsection{Description of the miniature heaters}
\label{sec:heaters}

\subsubsection{The wire}

The wire heater is made of a 90\% platinum -- 10\% rhodium alloy. It is
manufactured from a Wollaston wire by etching its
$\SI{50}{\mu m}$-diameter silver cladding. The wire diameter, as documented
by the manufacturer, is $d_w = \SI{1.3}{\mu m}$ and its length is
estimated from
resistance measurements to be $\SI{450}{\mu m}$.
It is essentially built the same way as it was in \textcite{Duri15} and
the main difference is that the present wire is soldered on a DANTEC
55P01 hot-wire support.

The resistivity of the Pt-Rh alloy decreases almost linearly with the
temperature from \SI{300}{K} down to 40 -- 50~{K}, and the sensitivity,
$dR_w/dT$, where $R_w$ is the wire's resistance, is therefore
almost constant.
Below this temperature, the sensitivity starts to decrease until it
eventually vanishes around \SI{13}{K}. For this reason it is
necessary to maintain the wire at temperatures well above \SI{13}{K},
in order to have access to its temperature through the resistance
measurement. We typically overheat it to $T_w \approx \SI{25}{K}$
which corresponds $R_w \approx \SI{36}{\Omega}$.

The wire heater is driven at constant resistance and thus at constant
temperature.

\subsubsection{The film}
\label{sec:per2}

\begin{figure}[!ht]
  \centering
  \includegraphics[width=1.0\linewidth]{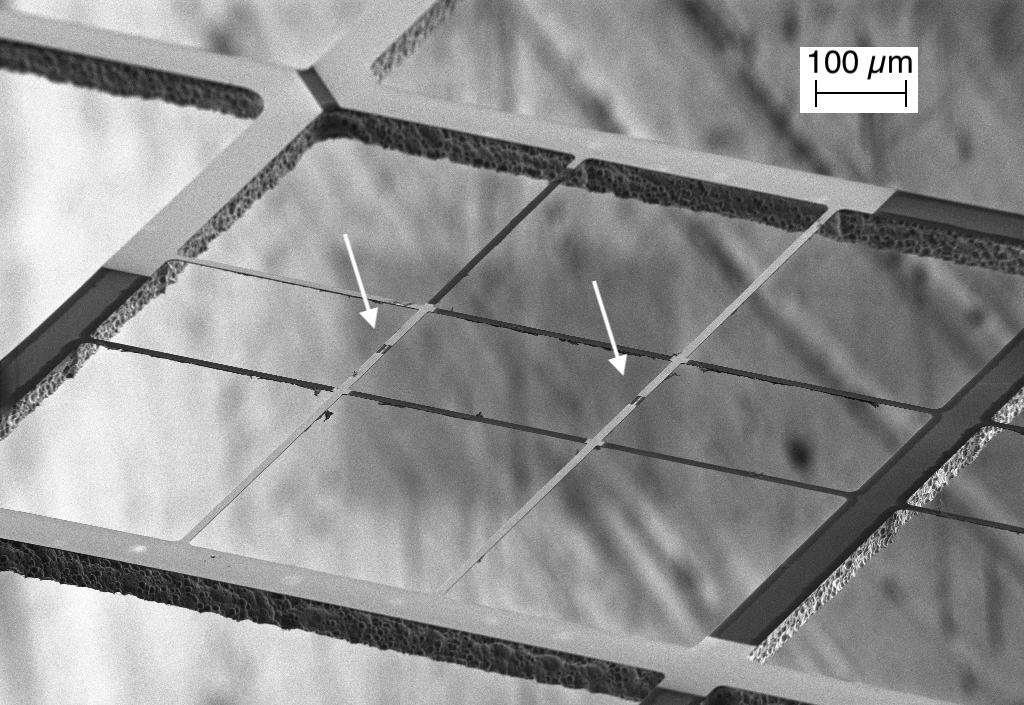}
  \caption{Electron microscope picture of the frame holding the film heater array.
  Two heating Pt strips are dark areas, pointed by white arrows, near the center
  of the supporting 1-mm-long SiN ribbons. A gold layer deposited on both sides of Pt
  provides the electrical contacts (lighter area). Thermal contact between Au and
  the Pt strip is reduced thanks to an intermediate buffer of Pt.}
  \label{fig:picper2}
\end{figure}

The film heater consists of a platinum thin-film strip, patterned
within a $\SI{2.6}{\mu m} \times \SI{5}{\mu m}$ area, and deposited on a
\SI{500}{nm}-thick, \SI{10}{\mu m}-wide and \SI{1}{mm}-long SiN ribbon
(see Fig.~\ref{fig:picper2}). The current leads to the Pt strip consist
of \SI{200}{nm} gold layers. As previously for the Pt-Rh alloy of the
wire heater, the temperature sensitivity of Pt electrical resistivity
vanishes around \SI{13}{K}~\cite{PlatiniumITS90}. In practice, the
heater is overheated up to few tens of Kelvins to benefit from a
nearly temperature-independent sensitivity.  Without overheating, the
resistance of the film is 730~$\Omega$ below \SI{10}{K} and
1060~$\Omega$ at \SI{77}{K}. To reach an overheating of \SI{25}{K} in
a quiescent \SI{2}{K} He~II bath, a current of $\SI{300}{\mu A}$ is
needed. Details about the microfabrication process of this 
probe will be provided in another paper.

This heating film is mounted in the grid flow -- with the film facing  upstream --
and driven with a constant-current electronics.

\subsubsection{The chip}
\label{sec:chip}

The chip heater is a bare Cernox\textsuperscript{\textregistered} CX-BR
thermometer from Lake Shore cryotronics Inc., mounted in the SHREK
experiment.
It consists of a $\SI{0.3}{\mu m}$ thick zirconium oxynitride film deposited
on a sapphire substrate whose dimensions are $\SI{0.2}{mm}\times
\SI{0.97}{mm}\times \SI{0.76}{mm}$~\footnote{Specification of bare chip Cernox thermometers (visited 12/2020):
\url{https://www.lakeshore.com/products/categories/specification/temperature-products/cryogenic-temperature-sensors/cernox}}.

Like semiconductors, and contrary to Pt-Rh and Pt heaters, the
resistance increases as the temperature decreases.
The sensitivity  $(T/R)dR/dT$ remains 
almost constant ($-0.45\pm 0.05$) over the explored
temperature range, from $\SI{1.7}{K}$ to \SI{30}{K}.
This contrasts with the two previous heaters which lose temperature
sensitivity below roughly \SI{13}{K}.

The probe is driven at slowly varying sinusoidal current $i(t)$:
\begin{equation}
 \label{eq:current}
 i(t) = I_0\sin \left(2\pi \frac{t}{\tau}\right),
\end{equation}
where $I_0$ is the current amplitude and $\tau$ is the period.
The resulting voltage $e$  across the chip together with the current
are recorded using a  NI-PCI-4462 acquisition board.

The period $\tau$ is typically 0.2~s, much larger than the thermal time
constant of the chip and than the turnover time of large eddies in the
flow. This allows us to determine continuously the temperature  of the
chip as a function of the input power, from bath temperature to
around \SI{30}{K}.

\section{Metastable heat transfer states at low temperature}
\label{sec:bistable}

At the lowest temperature explored in this study, $T_\infty =
\SI{1.74}{K}$ which corresponds to a superfluid fraction of 71\%, we
report  the observation of two metastable heat-transfer regimes. As
the external velocity over the wire heater increases, the less
conductive regime takes precedence over the more conductive one, in
terms of residence time in each metastable state.

This effect manifests itself as a \textit{decrease} of the averaged
heat transfer as velocity increases, at least 
in the intermediate range of velocity where both co-exist.
Rather than focusing on the average heat transfer,  this
effect is better illustrated by the histograms of the instantaneous
heat transfer.

\begin{figure}[!ht]
  \centering
  \includegraphics[]{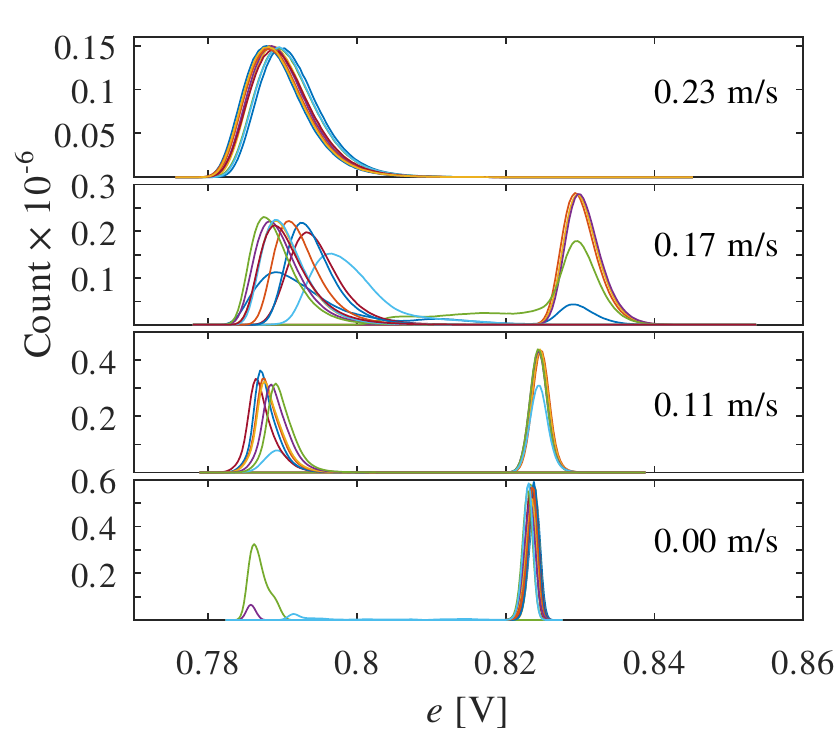}
  \caption{Histogram of the wire heater voltage output at
    $T_\infty = \SI{1.74}{K}$ for  various velocities. Each curve
    represents a dataset with  $4\times10^6$ samples.}
  \label{fig:histo_175}
\end{figure}

Figure~\ref{fig:histo_175} presents the histogram of the wire heater
voltage at the four smallest velocities. Each curve is the histogram
for one dataset, an approximately two-minutes-long segment of signal.
This duration is much  longer than the longest characteristic time
scales of turbulence at the heater location; these time scales are
of the order of only a fraction of a second (typically $M/v_\infty \lesssim
\SI{4}{mm}/\SI{0.1}{m.s^{-1}} \approx \SI{0.04}{s}$). In this regard, a 
segment of any signal's segment that belongs to one of the conduction
states can be considered quasistationary as far as
hydrodynamic phenomena are concerned, and the corresponding states can
be considered as stable or metastable. It cannot be fully excluded, though,
that the switching  from one state to the other is triggered by very rare
events in the flow.

At null velocity, the more conductive state is clearly the most probable and as
the velocity is increased the probability of observing this state progressively
decreases and eventually vanishes. In the present conditions, the difference in
heat transfer efficiency between the two states is around 10\% and both states
co-exist for $v_\infty \lesssim \SI{0.2}{m/s}$. Analysis of the time
series (not shown here) shows that the typical lifetime of each state
is of the order of tens of seconds. For this reason, the two-state
behavior described here should not be 
confused with that described below in Sec.~\ref{sec:pulsetime} for the
film heater signal. In the latter case, no metastable behavior
will be observed: The persistence time of the most conductive state
will be typically four to five decades shorter, and of the order of the
shortest resolved time scale of the turbulence.

In the following section, which addresses the mean heat transfer
versus mean velocity, the velocity response of each state will be
examined separately.

\section{Effect of the velocity on the mean heat transfer}
\label{sec:velocityeffect}
In this section we analyze the sensitivity of the mean heat transfer
to the velocity of the surrounding flow. Using the chip heater, 
we first show that the sensitivity is conditioned to the presence of an He~I
film at the surface of the heater. Then we use the wire heater to
determine how the temperature of the surrounding He~II affects the sensitivity
to the velocity.

\subsection{Sensitivity to velocity conditioned to the  presence of an
  He~I film}

We report here that the heat transfer from a heater immersed in He~II
becomes velocity dependent concomitantly with the formation of an He~I
film around the heater.

\begin{figure}[!ht]
  \centering
  \includegraphics[]{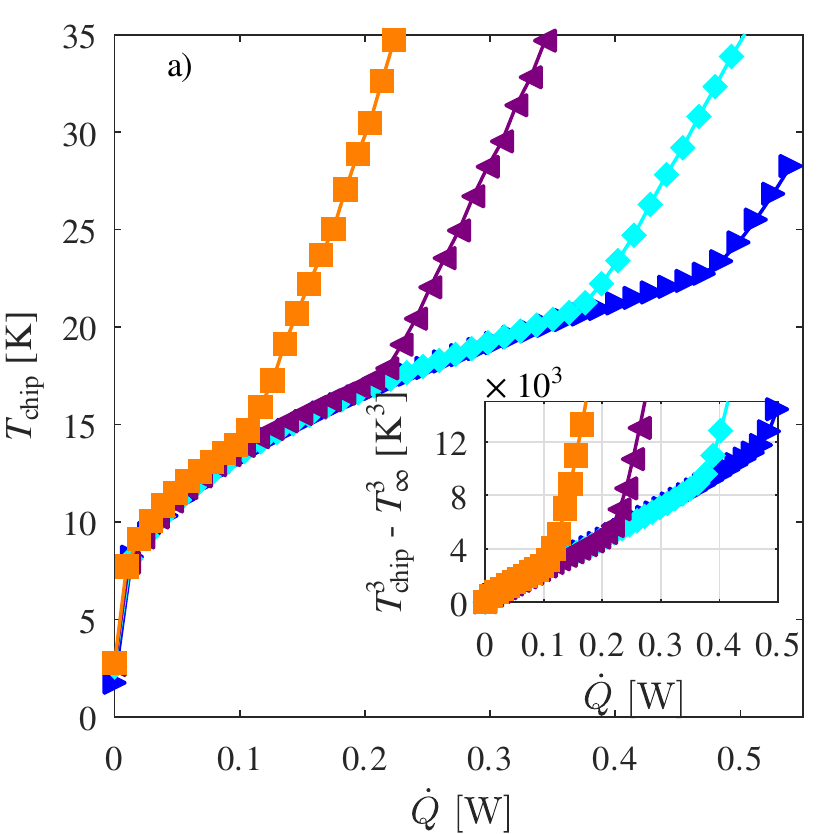}
  \includegraphics[]{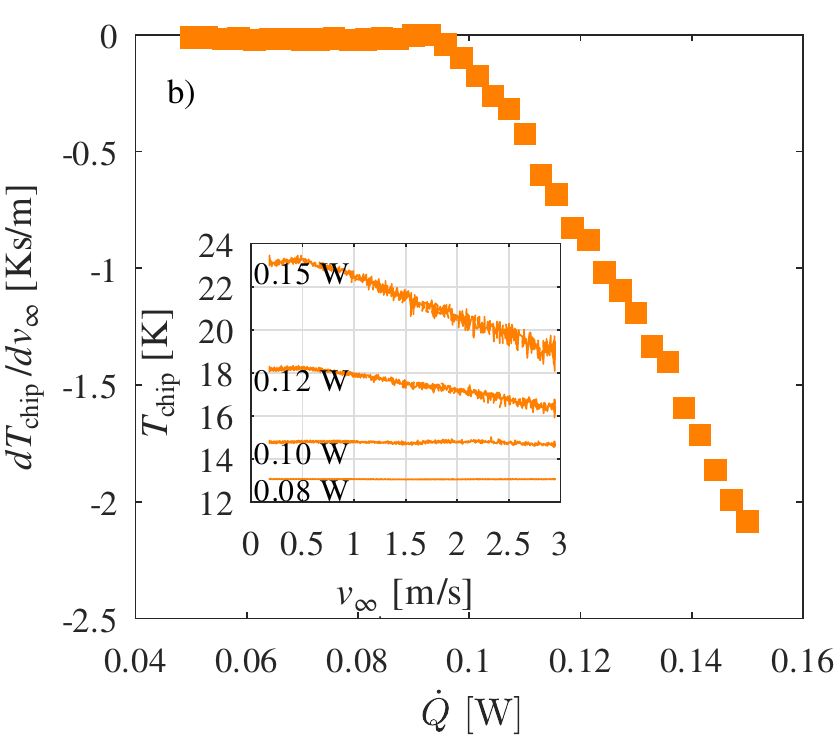}
  \caption{a) Temperature $T_\text{chip}$ of a bare chip
    Cernox\textsuperscript{\textregistered}
    as a function of the dissipated electrical power for various bath
    temperatures:
    {\color{orange}$\blacksquare$} $\SI{2.11}{K}$,
    {\color{violet}$\blacktriangleleft$} $\SI{2.05}{K}$,
    {\color{cyan}$\blacklozenge$} $\SI{1.93}{K}$,
    {\color{blue}$\blacktriangleright$} $\SI{1.74}{K}$. Inset:
    $T_\text{chip}^3-T_\infty^3$ as a function  of $\dot Q$.\\
    b) Average sensitivity of the temperature of the chip to the velocity in the
    range $1-\SI{1.5}{m/s}$ as a function of the input power.
    Inset: Temperature of the chip as a function of the velocity for
    various dissipated electrical powers. The temperature of the bath is
    $T_\infty = \SI{2.11}{K}$, corresponding to the orange curve in panel (a).}
  \label{fig:cernox}
\end{figure}

Figure~\ref{fig:cernox}(a) shows the power
required to overheat the chip heater in the
absence of an external flow.

As expected, at the lowest power input, below approximately $\SI{10}{\mu W}$, 
the chip temperature $T_\text{chip}$ is close to
the bath temperature $T_\infty$. This part of the curve
is not detailed.

For power inputs larger than $\SI{10}{\mu W}$, the chip temperature is
measurably larger than the bath temperature. At intermediate power
inputs exceeding $\SI{10}{\mu W}$ the curves
for all bath temperatures tend to collapse on a single
baseline curve, but for larger power inputs,
above a bath temperature dependent critical power $\dot
Q_\text{crit}$, the chip temperature starts to increase with $\dot Q$ 
much more rapidly.

As illustrated by the inset of Fig.~\ref{fig:cernox}(a), the temperature
baseline common for of all curves in the intermediate power range evolves 
roughly as $T_\text{chip}^n-T_\infty^n\propto \dot Q$ with $n=3$.
Such a dependence is typical of a heat transfer limited mostly by a
large-heat-flux Kapitza resistance. For instance Van Sciver
compilation reports  exponents of $n=3\pm0.5$ (see p.~293 in
Ref.~\cite{VanSciverLivre2012}). This thermal resistance appears at the
interface between the chip and helium, and at the inner solid
interfaces within the chip. It is responsible for a significant
overheating of the chip  ($T_\text{chip}$) compared to the liquid
helium in contact with it ($T_\text{chip}^\prime$)
\footnote{In principle, in the relation 
  $T_\text{chip}^n-T_\infty^n\propto \dot Q$
  the temperature of helium at the chip
  interface, $T_\text{chip}^\prime$ should be used instead of $T_\infty$.
  However, such a
  correction would hardly alter the overlapping of the baselines since
  $T_\text{chip}^\prime - T_\infty < T_\lambda - T_\infty \ll
  T_\text{chip}$ for (large) heat flux not exceeding $\dot
  Q_\text{crit}$}.

With this in mind, the critical heat flux $\dot Q_\text{crit}$ is
interpreted as the threshold at which the temperature
$T_\text{chip}^\prime$ of helium at the solid-liquid interface becomes larger than
$T_\lambda$. Above this threshold, a thin He~I layer forms around the
heater. Since He~I  is significantly less conductive than He~II,  the
chip temperature grows much more rapidly as the heat flux is increased
beyond $\dot Q_\text{crit}$. A similar phenomenology is reported in
the ``film boiling'' literature when a heater is overheated in a bath of
He~II at saturated vapor pressure, instead of a bath of pressurized helium in
our case. In this case, a helium gas layer forms around the heater and
also contributes to thermal isolation of the heater from its surrounding.

Figure~\ref{fig:cernox}(b) presents an important result. In the inset,
the heater's mean temperature is displayed versus the mean velocity of
the surrounding flow at $T_\infty = \SI{2.11}{K}$ (20\% superfluid
fraction). In the main axes, the average sensitivity $d
T_\text{chip}/d v_\infty$ in the velocity range from 1 to 1.5~m/s is
displayed as a function of the input power.
At the lowest heater power, no  velocity dependence is discernible. This
absence of sensitivity is observed down to 1.74~K, the lowest tested
bath temperature, and is consistent with the
standard understanding  of heat transfer in He~II~\cite{Johnson78}.
Above $\dot Q \approx \SI{0.093}{W}\approx \dot Q_\text{crit}$, some
sensitivity  starts  to develop.
In other words, the observed velocity sensitivity is concomitant with
the appearance of the He~I layer surrounding the heater.
As the power increases, the He~I layer is expected to thicken thus
leading to an increase, observed in our experiment, of the magnitude
of sensitivity.

\subsection{Velocity-Temperature dependence of heat transfer }

The chip heater, described above in Sec.~III, is not well-suited to
explore experimentally the basic mechanism of forced heat transfer. First, due
to its ``large'' size and the sharp angles of its parallelepiped
shape, its wake
is highly turbulent at all velocities, which complicates
modeling. Second, it is assembled with different materials leading to
a larger Kapitza resistance and larger temperature inhomogeneity
within the heater and thus at its surface. Third, its shape does not
have any simple symmetry which could ease analytical description of
heat transfer. Other limitations arise from the flow facility as 
it is not optimized to produce low
velocity and thus a less turbulent wake on the heater. Besides, the
velocity field in the vicinity of the heater is poorly known.

For all these reasons,  systematic measurements have been performed 
in the grid flow using the wire heater. In these conditions the flow
of He~I over the wire can be regarded as laminar: its characteristic
Reynolds number $\text{Re}=d_w v_\infty / \nu$, with $d_w=\SI{1.3}{\mu m}$,
$v_\infty=\SI{0.2}{m/s}$, and $\nu=\SI{2e-8}{m^2.s^{-1}}$, is of the
order of 10. In contrast, the corresponding Reynolds number of the
flow around the chip heater is three decades larger, well beyond wake
instability thresholds.

\begin{figure}[!ht]
  \centering
  \includegraphics[]{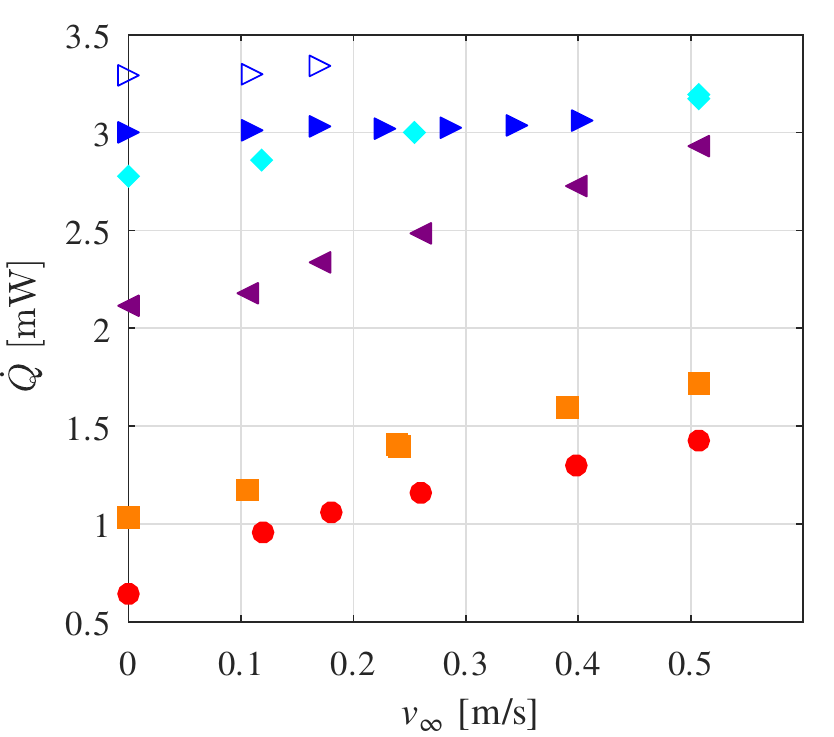}
  \caption{Electrical power required to regulate at \SI{25}{K} the
    wire heater as
    a function of the flow mean velocity for various bath
    temperatures: {\color{red}$\bullet$} $\SI{2.28}{K}$ (0\% superfluid),
    {\color{orange}$\blacksquare$} $\SI{2.13}{K}$ (10\% superfluid),
    {\color{violet}$\blacktriangleleft$} $\SI{2.05}{K}$ (31\% superfluid),
    {\color{cyan}$\blacklozenge$} $\SI{1.93}{K}$ (51\% superfluid),
    {\color{blue}$\blacktriangleright$} $\SI{1.74}{K}$ (71\%
    superfluid) in the ``less conductive'' regime, see
    Sec.~\ref{sec:bistable}),  {\color{blue}$\triangleright$}
    $\SI{1.74}{K}$ (71\% superfluid) in the ``more conductive'' regime.
    Solid lines indicate the best linear fit for each data series.}
  \label{fig:calibHeII}
\end{figure}

Figure~\ref{fig:calibHeII}  presents the electrical power
required to regulate the wire heater at \SI{25}{K}
versus the mean velocity, for flow temperatures ranging
between $\SI{1.74}{K}$ (71\% superfluid fraction) and $\SI{2.28}{K}$ (0\%
superfluid fraction).
It shows that when the heater is submerged into an external flow, an
additional electrical power is required to maintain its temperature.
This conclusion is consistent with the previous observation in a jet
flow~\cite{Duri15} but we can now  resolve more precisely the bath temperature
dependence of $\dot{Q}$.

\begin{figure}[!ht]
  \centering
  \includegraphics[]{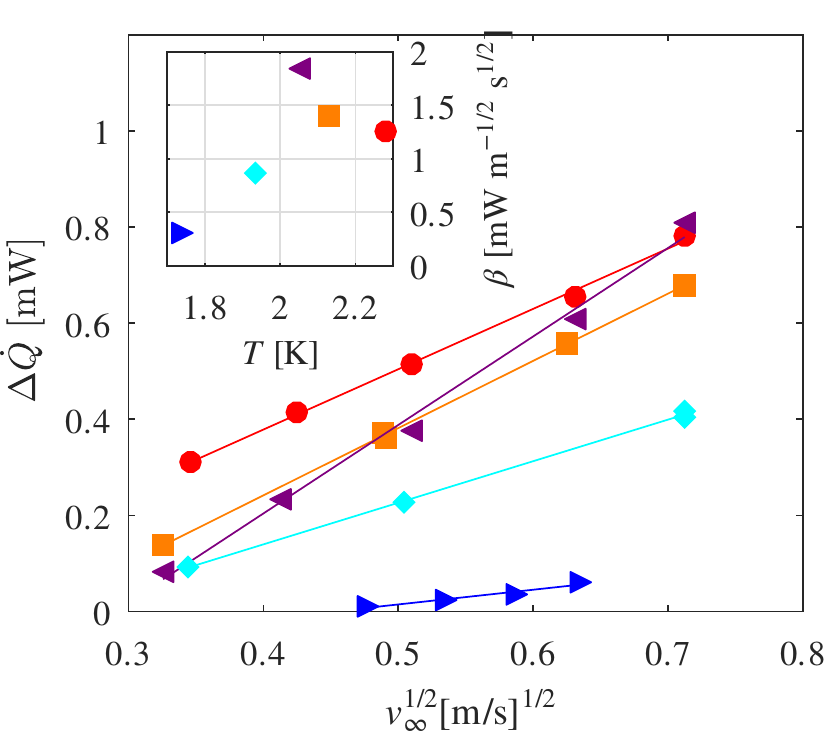}
  \caption{Time average of the excess power required to overheat the
    wire at \SI{25}{K} once the flow is turned on versus the square root
    of the velocity for various bath temperatures:
    {\color{red}$\bullet$} $\SI{2.28}{K}$,
    {\color{orange}$\blacksquare$} $\SI{2.13}{K}$,
    {\color{violet}$\blacktriangleleft$} $\SI{2.05}{K}$,
    {\color{cyan}$\blacklozenge$} $\SI{1.93}{K}$,
    {\color{blue}$\blacktriangleright$} $\SI{1.74}{K}$ in the ``less
    conductive'' regime (see
    Sec.~\ref{sec:bistable}). The markers indicate the
    actual computed values while the lines show the best linear
    fit of the data corresponding to Eq.~(\ref{eq:fit_king}).
    Inset: values of the slopes $\beta$ for all
    temperatures.}
  \label{fig:calibHeII_pow}
\end{figure}

In two-dimensional laminar flows (such as, e.g., the flow
around a thin wire) of classical fluids, at high P\'eclet numbers
$\text{Pe}=\text{Re}\cdot\text{Pr}$, where $\text{Pr}=\nu/D\gg1$ is the
Prandtl number, with $D$ being the fluid thermal diffusivity, the heat transfer
rate between a solid surface, and the fluid scales as $v_\infty^{1/2}$.
This follows from the analysis~\cite{Lighthill50,Acrivos60} of the
convective-diffusive heat transfer in the thermal boundary layer. For the
forced  heat transfer around a heated wire this scaling has been experimentally
and empirically confirmed in, e.g., Ref.~\cite{Collis59}.
In Fig.~\ref{fig:calibHeII_pow} we thus present the excess power
\begin{equation}
  \Delta \dot Q (T_\infty , v_\infty)=\dot Q(T_\infty , v_\infty)-
  \dot Q(T_\infty , 0 )
\end{equation}
required to maintain the temperature of the wire at \SI{25}{K} once
the flow is turned on,  as a function of the square root of the velocity.

The best fit of the form
\begin{equation}
  \label{eq:fit_king}
  \dot Q (T_\infty , v_\infty ) =
  \zeta(T_\infty) + \beta(T_\infty) \cdot v_\infty ^{1/2}
\end{equation}
is calculated omitting the data at null velocity as is customary in
standard fluids  where natural convection prevents the $v^{1/2}$ scaling to
hold down to small velocities. The coefficients $\zeta$ and $\beta$
are reported in Table~\ref{tab:q}.

For completeness, we also reported in Table~\ref{tab:q} the coefficients for 
a linear fit of the form
\begin{equation}
  \dot Q(T_\infty, v_\infty) = \chi(T_\infty) + \gamma(T_\infty) v_\infty.
  \label{eq:fit-linear}
\end{equation}
where $\chi$ and $\gamma$ are temperature-dependent coefficients.

Due to the limited range of velocities, the above fits do not allow us
to determine which of the two scaling laws, (\ref{eq:fit_king}) or
(\ref{eq:fit-linear}), is the best suited. At \SI{2.28}{K}, in He~I,
we know from experience the $v^{1/2}$ scaling is better suited, and
this probably remains true at \SI{2.13}{K}, but at all other
temperatures both laws could work.

A notable result, highlighted in the inset of
Fig.~\ref{fig:calibHeII_pow}, is the nonmonotonic dependence of the
sensitivity to velocity versus the superfluid fraction (or fluid
temperature $T_\infty$) with a maximum sensitivity somewhere between
superfluid fraction of 10\% and 50\%; also note that the sensitivity to velocity
significantly decreases for large superfluid fractions.

One point is worth stressing for subsequent modeling. For flow
temperatures $T_\infty \geqslant \SI{1.93}{K}$,
the  sensitivity to velocity, say defined as $d\dot Q/d v_\infty$,
varies only slightly with the temperature, while $\dot Q$
significantly depends on it. In particular, the sensitivity in high
temperature He~II is close to sensitivity in He~I, that is in the
absence of superfluid.

\begin{table}[ht!]
  \begin{tabular}{|l|c|c|c|c|c|}
    \hline
    $T_\infty$~[K]&2.28&2.13&2.05&1.93&1.74\\ \hline
    $X_{xf}[\%]$&0&10&31&51&71\\\hline
    $\dot Q(v_\infty=0)$ [mW] &0.64&1.03&2.12&2.78&3.00\\\hline
    \multicolumn{6}{|c|}{$\dot Q = \chi + \gamma \cdot v_\infty$}\\\hline
    $\chi\,\si{\left[mW\right]}$&0.74&1.05&2.06&2.78&3.00\\\hline
    $\gamma\,\si{\left[mW.m^{-1}.s\right]}$&1.45&1.37&1.68&0.81&0.12\\\hline
    \multicolumn{6}{|c|}{$\dot Q = \zeta + \beta \cdot v_\infty^{1/2}$}\\\hline
    $\zeta$ [mW]  &0.52&0.71&1.59&2.57&2.86\\\hline
    $\beta\,\si{\left[mW\,m^{-1/2}s^{1/2}\right]}$
                  &1.26&1.40&1.84&0.87&0.30\\\hline
  \end{tabular}
  \caption{Summary of the parameters obtained when fitting the
    power $\dot Q$ against $v_\infty$ (see Fig.~\ref{fig:calibHeII})
    or $v_\infty^{1/2}$. The wire heater is overheated at constant
    temperature, here \SI{25}{K}.}
  \label{tab:q}
\end{table}

The sensitivity to velocity versus the wire heater temperature was not 
explored, but the experiment with the chip heater indicates that it
can be significant  (see e.g. Fig.~\ref{fig:cernox}).

\section {High frequency peak: The Cooling glitches}
\label{sec:coolingglitches}

We now report  a puzzling feature of heat transfer in a quantum flow:
A well defined spectral peak in the PSD which we show can be attributed to
the quasiperiodic occurrence of intense short-lived heat flux
enhancements. These events have been named ``cooling glitches''.

\subsection{Emergence of a spectral peak}
\label{sec:coolingglitchesspectra}

\begin{figure}[ht!]
  \centering
  \includegraphics[]{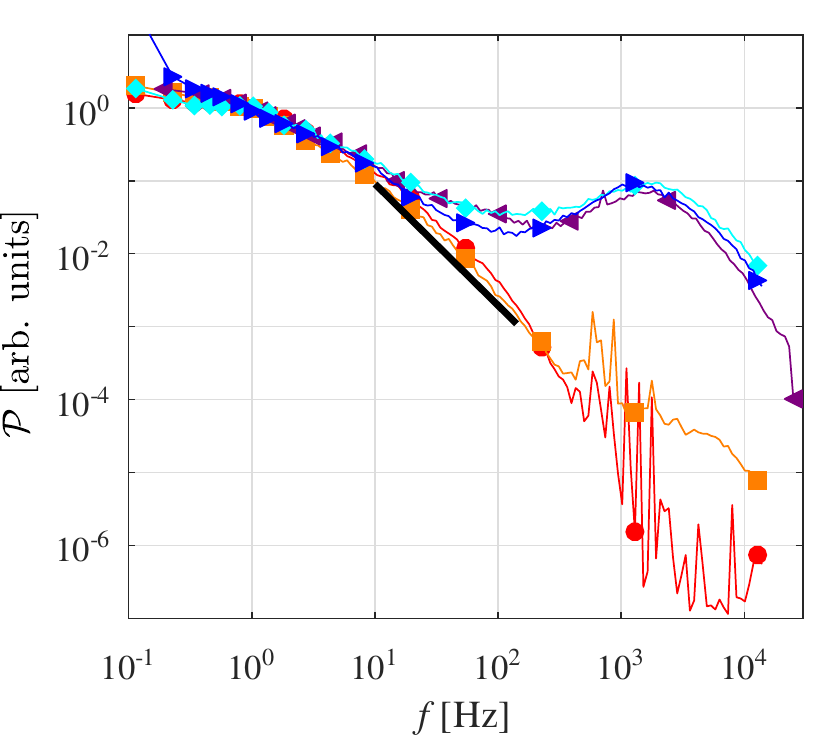}
  \caption{Power spectral density $\mathcal{P}(f)$ of the current in the
    wire at \SI{25}{K} for various flow temperatures:
    {\color{red}$\bullet$} $\SI{2.28}{K}$ (0\% superfluid),
    {\color{orange}$\blacksquare$} $\SI{2.13}{K}$ (10\% superfluid),
    {\color{violet}$\blacktriangleleft$} $\SI{2.05}{K}$ (31\% superfluid),
    {\color{cyan}$\blacklozenge$} $\SI{1.93}{K}$ (51\% superfluid),
    {\color{blue}$\blacktriangleright$} $\SI{1.74}{K}$ (71\% superfluid) in
    the ``less conductive'' regime (see
    Sec.~\ref{sec:bistable}). The black line shows a $f^{-5/3}$
    power law. In each case, the mean velocity
    is $0.250\pm \SI{0.015}{m/s}$. The amplitude of the signal is
    rescaled so that spectra overlap at $f = \SI{1}{Hz}$.}
  \label{fig:spectraTvar}
\end{figure}

The wire heater was inserted in the grid flow and its  temperature 
was maintained around \SI{25}{K}. The time series of the electrical 
current has been analyzed.

Figure~\ref{fig:spectraTvar} shows the power spectral density (PSD)
$\mathcal{P}(f)$ of the current in the wire for a superfluid fraction
varied from 0\% ($\SI{2.28}{K}$) up to 71\% ($\SI{1.74}{K}$) and at a
mean velocity $v_\infty = 0.250\pm \SI{0.015}{m/s}$.

In the absence of superfluid, the measured spectrum in the range of
intermediate frequencies is compatible with
the Kolmogorov spectrum of classical turbulence, as expected for grid
turbulence (see Appendix for further discussion). For
a 10\% superfluid fraction ($T_\infty =\SI{2.13}{K}$), the spectrum
departs from the Kolmogorov shape above $\approx \SI{500}{Hz}$. For superfluid
fractions equal to or larger than 31\% ($T_\infty\leqslant \SI{2.05}{K}$), a
broad spectral bump centered around $f_p \approx
\SI{1}{kHz}$ is observed. The bump is energetic enough to contribute
to most of the variance of the signal.

A departure from the classical turbulence spectra has  been previously
reported using a similar heated wire in a superfluid jet experiment
(see Fig.~2 in Ref.~[\onlinecite{Duri15}]), but the effect was much less
pronounced and no peak reported. A possible explanation for not
resolving a peak in this previous experiment is the combined effect of
insufficient time resolution (the maximum resolved spectral frequency
was \SI{5}{kHz}) and faster time scales of the jet flow. Indeed, compared to
the  conditions of Fig.~\ref{fig:spectraTvar}, the flow mean velocity
was five times larger and the variance of velocity fluctuations $48^2$
times larger (peak excluded), which could shift a possible peak beyond the maximum
resolved frequency.

\subsection{Evidences of cooling glitches}
\label{sec:coolingglitchesper2}

To gain more insight into the physical parameters that drive
the high frequency behavior, an additional experiment was  done using the
 film heater described in Sec.~\ref{sec:per2}.
 This heater was operated in the  grid flow at
$\SI{2.0}{K}$, but unfortunately it broke very rapidly so we
only have one velocity condition, $v_\infty=\SI{0.38}{m/s}$.

\begin{figure}[ht!]
  \includegraphics[]{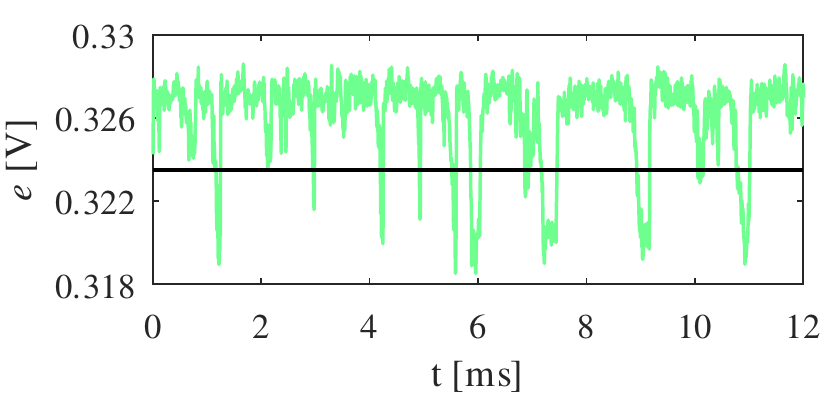}
  \caption{Sample of the film heater voltage at $T_\infty = \SI{2}{K}$ as a
    function of time, with a mean fluid velocity $v_\infty \approx
    \SI{0.38}{m/s}$ as a function of time.
    The black horizontal line marks the chosen threshold.}
  \label{fig:spot_signal}
\end{figure}

Figure~\ref{fig:spot_signal} shows a small portion of the signal from
the film heater. The heat transfer is enhanced during seemingly
random brief periods, lasting typically a tenth of a millisecond or
less.

\begin{figure}[ht!]
  \includegraphics[]{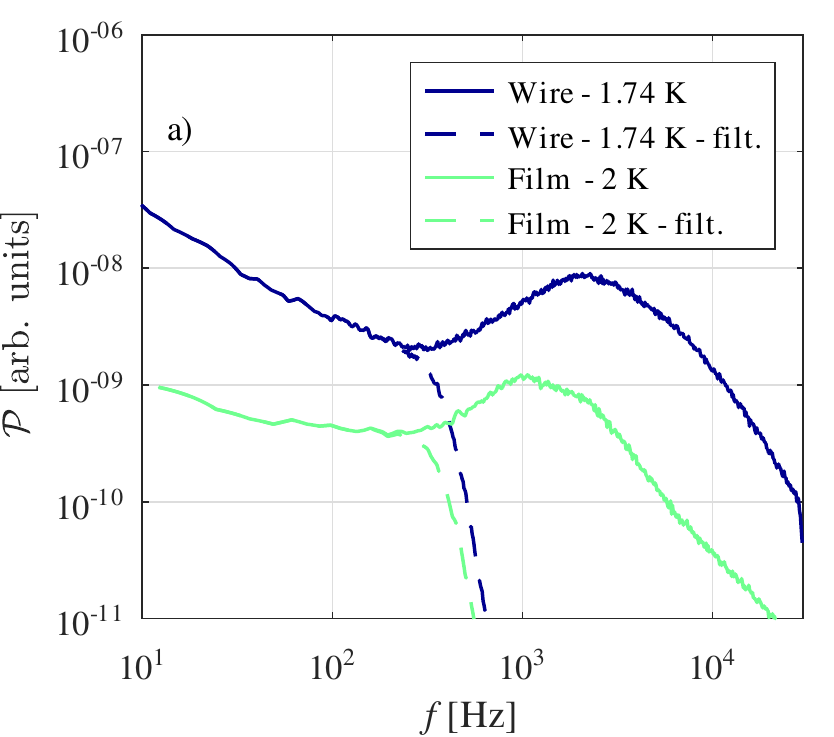}\\
  \includegraphics[]{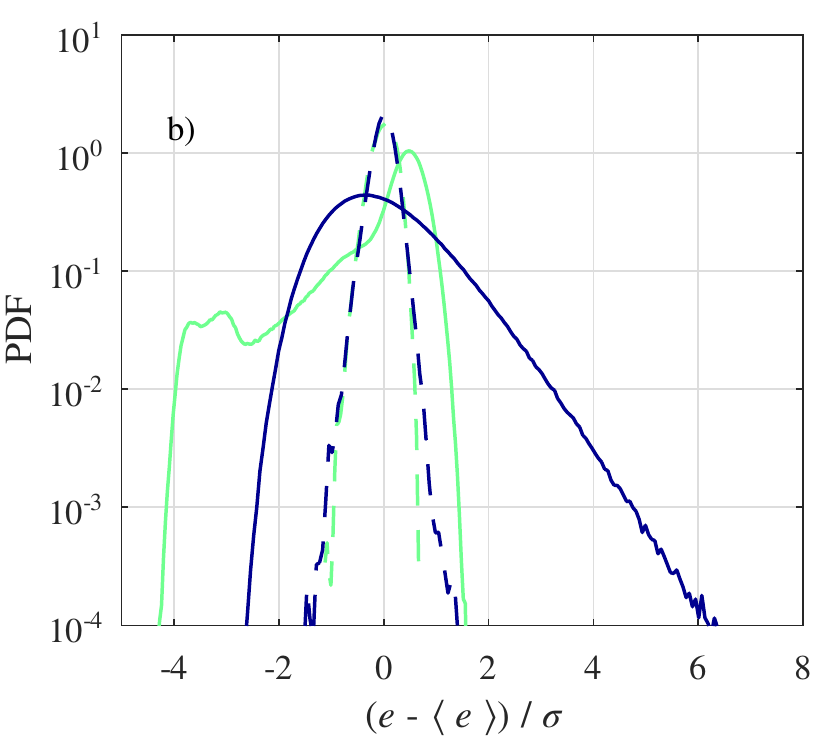}
  \caption{a) Comparison of the PSD, $\mathcal{P}(f)$, of the film heater
    at $T_\infty = \SI{2}{K}$ and of the wire heater at $T_\infty =
    \SI{1.74}{K}$, , with a mean fluid velocity $v_\infty
    \approx \SI{0.38}{m/s}$. Dashed lines correspond to the same data,
    but low-pass filtered at 400~Hz.
    b) Probability density function of the film and wire
    output signals in the same conditions as in a). The signals are
    centered and normalized by the
    standard deviations $\sigma$ of the unfiltered signals.}
  \label{fig:spot_vs_wire}
\end{figure}
The recorded time series for this smaller heater evidences the same 
spectral peaking  at high frequency as illustrated by
Fig.~\ref{fig:spot_vs_wire}(a) which displays the PSD,
$\mathcal{P}(f)$, from this film heater together with that from the
wire at the same velocity but lower temperature ($\SI{1.74}{K}$).
The bumps, even though they do not have the exact same shapes for the
film and the wire, are located at nearby frequencies. This rules out the length
of the heaters as a parameter governing the apparition of the bump since
they have very different length (by two orders of magnitude).
This also lets us assert that the electronic driving mode is not at
fault: Whether the heater is driven at constant temperature (wire) or constant
current (film), the bump remains.

Figure~\ref{fig:spot_vs_wire}(b) shows the centered and normalized
probability density function (hereafter PDF) of the output signal
recorded from the wire and the film electronic drivers.

A first observation is that the PDF of the signals are skewed in
opposite directions: The wire shows large excursions
towards high current (positive skew $s\approx 0.83$), while the film
shows large excursions towards low voltage (negative skew
$s\approx -2.0$).
These skewed PDF evidence that  both heaters record rare and intense
heat-flux events.
The opposite signs of the skewness are easily explained by the
difference in electronic drivers: The wire heater is driven at constant
temperature while the film is driven at constant current. An increase
of the cooling efficiency increases the current in the wire, but
decreases the temperature of the film, and thus the measured voltage
drop across it. Thus, both PDF indicate the existence of rare and
intense events of enhanced heat transfer between the heaters and the
flow. In the following, these events will be nicknamed ``cooling
glitches''.

The dashed lines in Fig.~\ref{fig:spot_vs_wire}(a) and
Fig.~\ref{fig:spot_vs_wire}(b) show respectively the PSDs and the PDFs of the
same signals after a  low-pass filtering at
$\SI{400}{Hz}$. As can be seen in Fig.~\ref{fig:spot_vs_wire}(a),
the result of the filtering is the suppression of the spectral bump,
while in Fig.~\ref{fig:spot_vs_wire}(b) we can see that each PDF becomes
almost gaussian. To be precise, both PDFs end up
with a small negative skew of order $s\approx-3.10^{-3}$, as
expected for standard hot-film and hot-wire anemometer in a turbulent
flow of low turbulent intensity. Indeed,  assuming that the PDF of the
velocity is gaussian,
the recorded PDF must be negatively skewed since the
sensitivity to the velocity decreases with velocity.
This filtering test strongly suggests that the cooling glitches, and
the broad frequency peaks refer to the same phenomenon. 

\label{sec:pulsetime}
\begin{figure}[!ht]
  \centering
  \includegraphics[]{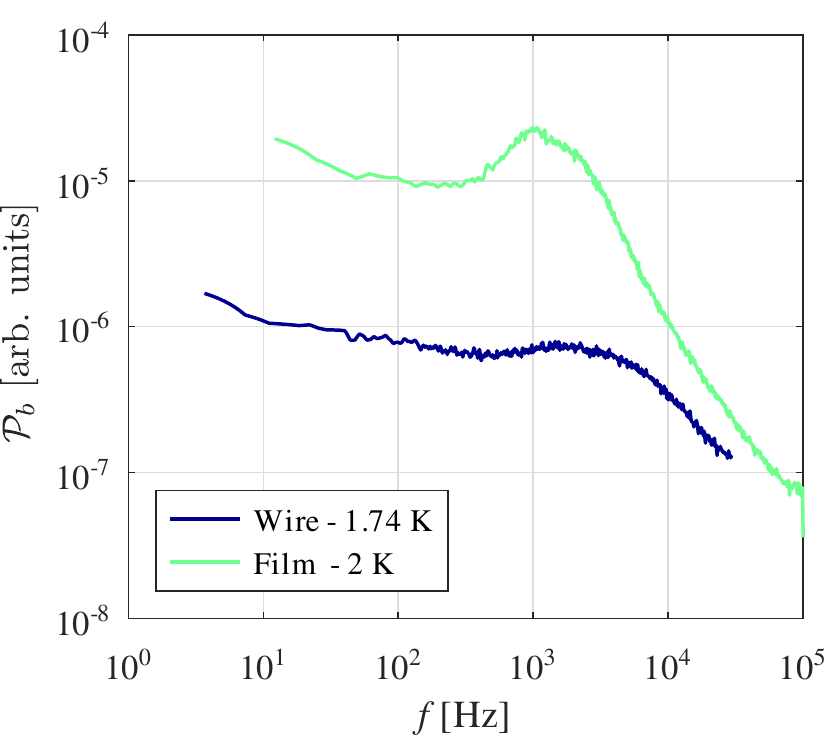}
  \caption{Power spectral density $\mathcal{P}_b$ of the wire and film
    heaters binarized signals (see the text for detail).}
  \label{fig:spot_pulses}
\end{figure}

 The  bimodal shape of the film's PDF in
Fig.~\ref{fig:spot_vs_wire}(b) supports the view that the system is
continuously switching between two well defined heat exchange
modes: the default one, and one with a higher cooling efficiency.

In the time domain, the occurrence of a cooling glitch on the film
heater can be spotted using an arbitrary threshold value,  for
instance the average between the peaks of both modes in the PDF
($e_\text{thresh}\approx 0.3235V$, see the black line in
Fig.~\ref{fig:spot_signal}.
On the other hand, the PDF from the wire time series does not allow
to resolve  two distinct modes, possibly because of a lower
temporal resolution. In order to binarize the wire heater signal we chose to define the
threshold value as $e_\text{thresh} = \langle e\rangle + 3\sigma$
where $\sigma$ is the standard deviation of the signal.

Figure~\ref{fig:spot_pulses} shows the PSD of the binarized signal,
$\mathcal{P}_b(f)$, for both the film and the wire heaters. For the
film, which has a clear bimodal behavior, the spectral bump is
preserved and the frequency of its maximum is  unchanged. The result
is essentially the same for the wire except that the bump is much less
pronounced than in the PSD of the raw signal.

The binarized signal only contains information about the
temporal distribution of gliches, i.e. their duration and the time interval
between them.
The fact that this very basic signal has a spectral bump similar to
that of the original signal, is another strong evidence that the
glitches are the root cause of the spectral bump.

\subsection{Glitch characteristic frequency versus velocity}
\label{sec:fglitch}

We have shown above that the sequence of cooling glitches exhibits a
characteristic frequency scale of a few kHz in present flow
conditions. We now characterize how this \textit{glitch peak frequency}
varies with the flow mean velocity.

The $\SI{1.74}{K}$ dataset from the wire is more specifically explored because
it allows the most accurate quantitative assessments.
Indeed,  the sensitivity of the mean (and low frequency) signal  to
the velocity is the lowest and most of the fluctuations of the energy
of the signal  are concentrated in the high frequency bump.
At low velocity, we only used data acquired during a period of time where
the wire was in the less conductive state since they prevail at most velocities (see Sec.~\ref{sec:bistable}).

\begin{figure}[!ht]
  \centering
  \includegraphics[]{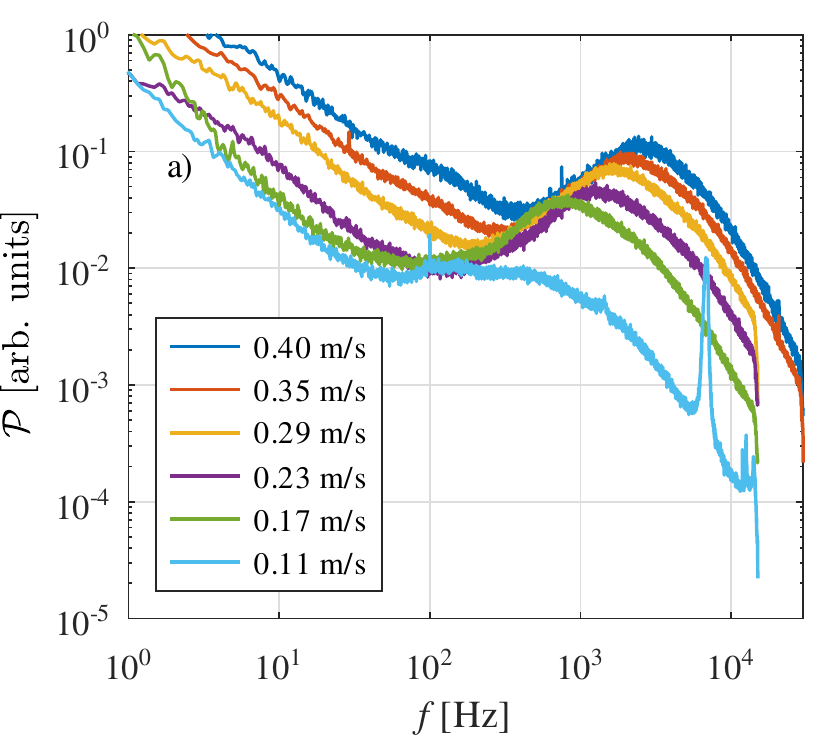}
  \includegraphics[]{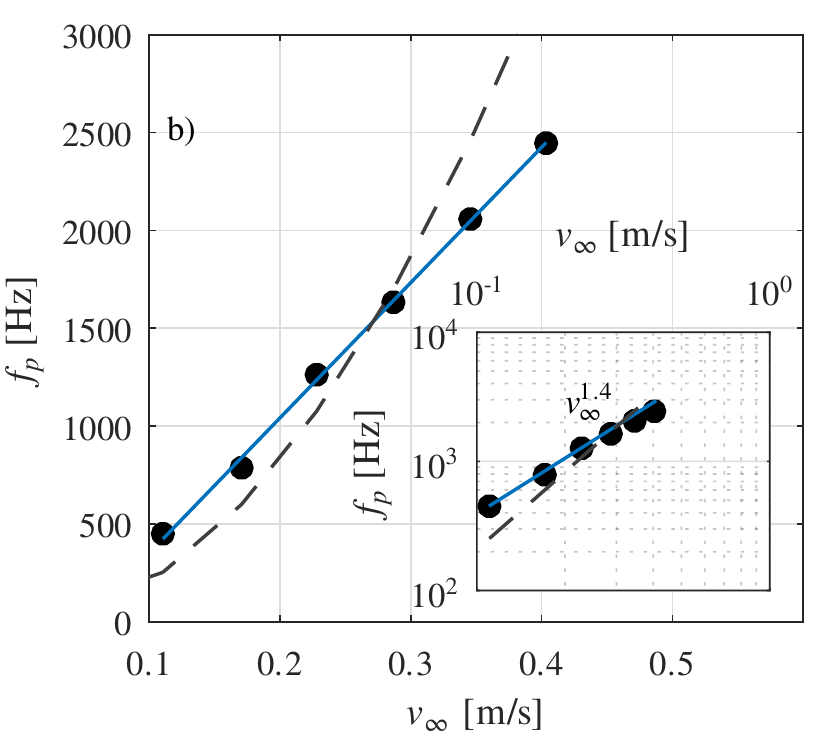}
  \caption{a) PSD $\mathcal{P}(f)$ of the hot-wire voltage at
    $\SI{1.74}{K}$ for various flow velocities.\\
    b) Frequency of the observed peak frequency as a function of
    the external flow average velocity ($\bullet$), together with a linear
    fit (solid blue line) and the result of the model developed in
    the next section, see Eq.~(\ref{eq:Strouhal}) (dashed line).\\
    Inset: Same data in log-log coordinates. Here the solid line is a
    fit with a power law $f_p\propto v_\infty^{1.4}$.}
  \label{fig:fpeak}
\end{figure}

Figure~\ref{fig:fpeak}(a) shows the power spectra of the wire heater
signal for various external flow velocities, and
Fig.~\ref{fig:fpeak}(b)  the evolution of the peak frequency versus
velocity. The peak  frequency, extracted using a local third-order
fit, is defined as the first local maximum above 500~Hz.
Over the explored range, the velocity dependence of the peak frequency
is consistent with an affine law $f_p =  a+b v_\infty^\alpha$ with
$\alpha \approx 1$, $a=\SI{-340}{Hz}$ and
$b=\SI{6912}{Hz.s.m^{-1}}$. This linear dependence
 suggests the existence of a fixed length scale in the flow of
order $1/b\approx \SI{150}{\mu m}$, i.e. much larger than the wire
diameter. The appearance of macroscopic length scales will be discussed in
Sec.~\ref{sec:wings}.

On a log-log scale, the best power law fit
of the data (see inset in
Fig.~\ref{fig:fpeak}(b)) is $f_p \sim v_\infty^\alpha$ with $\alpha \approx
1.4$.
Obviously, the limited range of velocity -slightly more than half a
decade- does not allow us to discriminate between both laws.

These scalings will be discussed in the next section.

\section{Discussion. Mathematical modeling of heat transfer in an He~II external flow}
\label{sec:discussion}
\subsection{Analytical model of heat transport at zero velocity}

At null velocity, \textcite{Duri15} showed that the mean heat
flux from a wire heater can be modeled satisfactorily
assuming that a thin supercritical He~I
layer surrounds the wire and concentrates most of the temperature gradient. In this
region, the temperature gradient $\bnabla T$ is proportional to
the heat flux $\bphi$, according to the standard Fourier
law: the fluid temperature decreases from  $T_w^\prime$ in He at the surface of the
wire ($r = r_w^+$), to $T_\lambda$ at $r=r_\lambda$. In the region
$r>r_\lambda$ the temperature gradient evolves
as~\cite{BonMardion79, Swanson85, Sato06}
\begin{equation}
  \label{eq:gm}
  \varphi ^ m = f(T) \frac{dT}{dr},
\end{equation}
where $f(T)$ is the so called conduction function; the power $m$ will be
specified later.

This basic model, solved numerically, enabled us~\cite{Duri15} to reasonably 
account for the mean heat transfer at all bath temperatures, including close
to $T_\lambda$.
In the following we solve the problem analytically.

Let $\dot Q$ be the heat rate needed to overheat the wire material at
a mean temperature $T_w$ in a liquid helium bath at temperature $T_\infty$.

The  problem is assumed to be axisymmetric and the  aspect-ratio of the wire large enough to neglect ends effect. In such
conditions, the wire temperature does not depend on the longitudinal
coordinate and the heat flux around the heated wire is given by

\begin{equation}
  \label{eq:phi}
  \varphi = \frac{\dot Q}{2\pi l r} = \frac{\Phi}{r},
\end{equation}
where $l$ is the length of the wire and $r (\ll l)$ is the radial
coordinate. Here the constant $\Phi$ is the heat transfer rate per radian and
per unit length.

Let $T_w^\prime$ be the temperature of helium in contact with the
wire. Due to the thermal resistance within the wire and Kapitza
resistance at the solid-fluid interface, $T_w^\prime < T_w$ and we can
define a thermal resistivity $\rho_K$ such that
\begin{equation}
  \label{eq:rK}
  T_w - T_w^\prime  =  \rho_K \Phi.
\end{equation}

This temperature difference is expected to be more significant for the
bulkier heaters (due to internal resistance), for nonmonolithic ones
(due to internal interface resistance), and at lower overheating (due
to larger Kapitza resistance at lower temperatures). For all reasons,
this temperature drop is expected to be more relevant for the chip
heater than for the wire heater. In the following, for simplicity, we
will simply refer to this temperature drop as the ``Kapitza
correction''.

In the supercritical He~I region, the Fourier law writes
\begin{equation}
  \label{eq:reg1}
  \frac{\Phi}{r} = - k \frac{dT}{dr},
\end{equation}
where $k$ is the thermal conductivity of helium. Neglecting the
temperature dependence of $k$, the integration of Eq.~(\ref{eq:reg1}) gives:

\begin{equation}
  \label{eq:reg1_int}
  \Phi \ln \left(\frac{r_\lambda}{r_w}\right) = k (T_w^\prime - T_\lambda).
\end{equation}

In the superfluid He~II region, Eq.~(\ref{eq:gm}) is integrated between
$T_\lambda$ (at $r=r_\lambda$) and $T_\infty$ (for $r \gg r_\lambda$):
\begin{equation}
  \label{eq:reg2_int}
  \frac{\Phi^m}{(m-1)r_\lambda^{m-1}} =
  \underbrace{\int_{T_\infty}^{T_\lambda}f(t) dT}_{F(T_\infty)}.
\end{equation}

Here we have introduced the conduction integral $F(T_\infty)$.
Eliminating $\Phi$ between Eqs.~(\ref{eq:reg1_int}) and (\ref{eq:reg2_int})
we obtain

\begin{equation}
  \label{eq:reg1_2}
  \ln \left(\frac{r_\lambda}{r_w}\right)=\frac{k (T_w^\prime - T_\lambda)}
  {\left[(m-1)r_\lambda^{m-1} F(T_\infty) \right]^{1/m}}\,.
\end{equation}

From the numerical solution~\cite{Duri15} of this problem, we know that the
width of the supercritical He~I layer, is small compared with the radius
of the wire (that is, $r_\lambda-r_w\ll r_w$), provided the bath temperature 
is not too close to $T_\lambda$ (say $T<2.1~\text{K}$). As
$\ln (r_\lambda/r_w) \approx (r_\lambda - r_w)/r_w\ll1$, this necessarily requires
that the right-hand side of Eq.~(\ref{eq:reg1_2}) is small. Introducing a small 
parameter
\begin{equation}
  \label{eq:epsilon}
  \epsilon=
    \frac{k (T_w^\prime -T_\lambda)}{\left[(m-1)r_w^{m-1} F(T_\infty)\right]^{1/m}}\ll1\,,
\end{equation}
and making use of the first-order asymptotic expansion of Eq.~(\ref{eq:reg2_int}) with 
respect to $\epsilon$, we obtain for the heat rate per radian and unit length:
\begin{equation}
  \label{eq:Phi-exp}
  \Phi(T_w^\prime ,\, T_\infty)=\Phi_{II}( T_\infty)\left[1+\epsilon\frac{m-1}{m}+O(\epsilon^2)\right]\,,
\end{equation}
where
\begin{equation}
  \Phi_{II}( T_\infty)  = \left[(m-1)r_w^{m-1} F(T_\infty)\right]^{1/m}.
\end{equation}
From Eq.~(\ref{eq:reg1_2}) it follows that the asymptotic expansion for $r_\lambda$,
which determines the width, $r_\lambda-r_w$ of the supercritical layer, should be
sought in the form
\begin{equation}
    \label{eq:r_lambda}
    r_\lambda=r_w(1+\epsilon+a_2\epsilon^2+...)\,.
\end{equation}
Making use of expansions (\ref{eq:Phi-exp}) and (\ref{eq:r_lambda}), Eq.~(\ref{eq:reg1_int})
can now be used to calculate the second-order term (i.e., the
coefficient $a_2$) of the expansion~(\ref{eq:r_lambda}). However, the
second (and higher) order corrections are of no interest in the
context of this work. 

Neglecting the corrections of order $\epsilon^2$ and higher and making use of 
Eq.~(\ref{eq:epsilon}), which can be written as $\epsilon\Phi_{II}=k (T_w^\prime -T_\lambda)$,
it is more convenient to represent relation~(\ref{eq:Phi-exp}) in the form
\begin{equation}
  \label{eq:PhiTot}
  \Phi(T_w^\prime ,\, T_\infty) \approx \Phi_I(T_w^\prime ) + \Phi_{II}( T_\infty),
\end{equation}
where
\begin{equation}
  \Phi_I(T_w^\prime ) = \frac{m-1}{m}k (T_w^\prime - T_\lambda)\,.
\end{equation}
Here the heat flux per radian and per unit length appears
as the sum of a contribution $\Phi_I(T_w')$ due to the conduction
in He~I and a bath temperature-dependent
contribution $\Phi_{II}(T_\infty)$ due to heat transport in He~II. 
As $\epsilon\ll1$, at low temperatures the former is much smaller than the latter.
It is worth noting that such an additive contribution of heat fluxes
is counterintuitive in thermal systems with resistances in series. 

Making use of Eq.~(\ref{eq:rK}), Eq.~(\ref{eq:PhiTot}) can be rewritten in
terms of $T_w$ (instead of $T_w^\prime$) and $T_\infty$:
expression versus $T_w$:
\begin{equation}
  \label{eq:PhiTot2}
  \Phi(T_w ,\, T_\infty) \approx \frac{1}{1+ K} \left[ \Phi_I(T_w )
    + \Phi_{II}( T_\infty)  \right],
\end{equation}
where
\begin{equation}
  \label{eq:Kapito}
  K = \rho_K \frac{m-1}{m}k
\end{equation}
is a dimensionless parameter, later referred to as the Kapitza
correction parameter which accounts for the strength of the temperature
drop between solid and liquid.

\begin{figure}[!ht]
  \centering
  \includegraphics[]{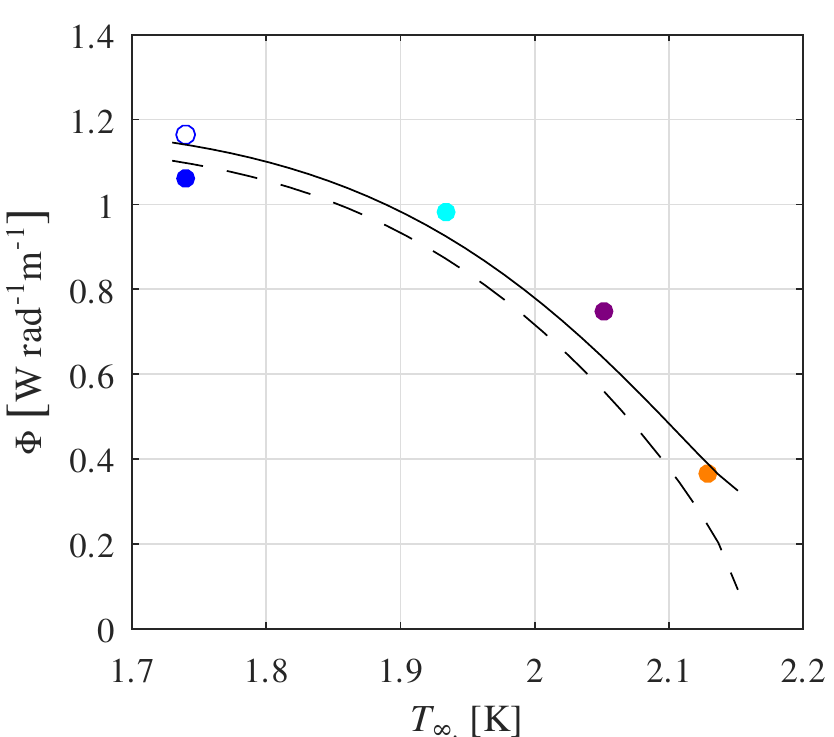}
  \caption{Comparison of the measured and the modeled heat flux 
    per radian and
    unit length (o). The solid line is the heat flux modeled according to
    Eq.~(\ref{eq:PhiTot}), with no adjustable parameter, and the dashed
    line shows the exact numerical solution. Computations were
    done using $T_w = \SI{25}{K}$.}
  \label{fig:null_vel_sato}
\end{figure}

Figure~\ref{fig:null_vel_sato} presents the measured and the simulated
values of the heat transfer rate $\Phi$ per unit length and radian 
as a function of temperature. We used the
Bon-Mardion/Sato~\cite{BonMardion79, Sato06} form of
Eq.~(\ref{eq:gm}), with $m = 3.4$. The thermal conductivity of
supercritical helium depends on temperature so we used its average
value $k = \SI{0.02}{W.m^{-1}.K^{-1}}$, determined using the
HEPAK\textsuperscript{\textregistered} library over the range
\SI{5}{K} -- \SI{25}{K}. Finally, the Kapitza correction was assumed
negligible for the wire ($K\ll 1)$.
As can be seen, the above simple model yields reasonably good
approximations for both
absolute values and the temperature dependence, without any adjustable
parameter.
The contribution of the He~I layer, $\Phi_I \approx
\SI{0.32}{W.rad^{-1}m^{-1}}$, is about 28\% the total heat flux
at \SI{1.74}{K} and 87\% at \SI{2.13}{K}.

We also solved Eq.~(\ref{eq:reg1_2}) numerically to estimate $r_\lambda$
and then computed the exact value of the heat flux from
Eq.~(\ref{eq:reg1_int}) (see dashed line in
Fig.~\ref{fig:null_vel_sato}). The relative error in the estimate of
the total heat flux is about 4\% at \SI{1.74}{K} and 40\% at
\SI{2.13}{K}. As expected, below \SI{2.1}{K} the linear
approximation $r_\lambda=r_w(1+\epsilon)$ [see Eq.~(\ref{eq:r_lambda})]
is quite reasonable.

\subsection{Analytical model of heat transport at finite velocity}
\label{sec:model_vit}
We now use an empirical approach to extend this analytical model and
account for the extra heat transfer observed in the presence of the
external flow.

The occurrence of one cooling glitch results in an increase of heat
transfer. Thus, in principle, the
overall velocity dependence of heat transfer could result from or be significantly
affected by a change in the statistics of occurrence of glitches or a change in
their strength. Still, this possibility could be discarded by the analysis of
the histograms of instantaneous heat transfer. Indeed, they reveal
that the most probable instantaneous heat transfer, which does not coincide with the occurrence of a glitch, has nearly the same velocity
dependence as the mean heat transfer, glitches included.

As pointed out earlier, the  sensitivity to velocity, say
$d\Phi/dv_\infty$ varies only slightly with the flow temperature
within the interval 
$\SI{1.93}{K} \leqslant T_\infty \leqslant \SI{2.28}{K}$, although $\Phi$,
or, more precisely, $\Phi_{II}$ vary significantly.
The (chip) heater, sensitive to temperature near $T_\lambda$,
has revealed that the velocity dependence is bound to the presence
of a He~I layer. The velocity dependence will thus be modeled by a modification
$\Phi_I^\star(T_w^\prime, \, T_\infty, \, v_\infty )$ of the contribution
$\Phi_{I}(T_w^\prime)$, so that
\begin{equation}
  \label{eq:toto1}
  \Phi(T_w , T_\infty, v_\infty) =
  \Phi_I^\star(T_w^\prime , T_\infty, v_\infty ) + \Phi_{II}( T_\infty).
\end{equation}

As customary in classical flows, the velocity dependence can be formally
embedded in the Nusselt number $\text{Nu}^\star(T_w , T_\infty, v_\infty)$ 
defined by the relation
\begin{equation}
  \label{eq:toto2}
  \Phi_I^\star(T_w , T_\infty, v_\infty) = \text{Nu}^\star \cdot \Phi_I(T_w^\prime).
\end{equation}

This definition of $\text{Nu}^\star$ is related to the classical Nusselt
number $\text{Nu}$ of the heat transfer from an arbitrary bluff body:
$\text{Nu}^\star(\text{Re})=\text{Nu}(\text{Re})/\text{Nu}(0)$.

As discussed above in Sec.~V [see Fig.~\ref{fig:calibHeII_pow} and
Eq.~(\ref{eq:fit_king}) in particular], the heat transfer rate
is consistent with a $v_\infty^{1/2}$ scaling with velocity,
provided the magnitude of the velocity is sufficiently away from $v_\infty=0$.
We, therefore, will adopt the following model for the Nusselt number:
\begin{equation}
  \label{eq:Nusselt}
  \text{Nu}^\star = A + B \text{Re}_{w}^{1/2},
\end{equation}

where $A$ and $B$ are dimensionless constants of the order unity and 
$\text{Re}_{w}$ is a Reynolds number based on the diameter of the wire
[the definition is given below,
see Eq.~(\ref{eq:Rew})].

In classical hydrodynamics, Eq.~(\ref{eq:Nusselt}) is known as King's
law and accounts for forced heat transfer from hot-wire anemometers (see
e.g. Ref.~[\onlinecite{Collis59}]). The square root dependence is understood as the signature
of the thermal  boundary layer around the anemometer.

The reason why $ \text{Nu}^\star(\text{Re}_{w}=0) =A \neq 1$
reflects the existence of an alternative heat transfer mechanisms at
zero velocity, e.g. natural convection.

Helium at $\SI{2.28}{K}$ is a classical fluid, and our wire heater
resembles a hot-wire anemometer. Hence, it is not surprising that
Eq.~(\ref{eq:Nusselt}) accounts for heat transfer measurement above the
superfluid transition at $T=T_\lambda$. The persisting agreement of 
Eq.~(\ref{eq:Nusselt})
in a superfluid bath strongly  suggests that a similar phenomenology
remains at play, in particular the stretching of the He~I layer
surrounding the heater by the incoming flow.

Thus, for $T<T_\lambda$ the Reynolds number $\text{Re}_{w}$ is  defined as
 \begin{equation}
  \label{eq:Rew}
  \text{Re}_{w}=\frac{d_w V_\text{eff}}{\nu},
\end{equation}
where $\nu$ is the kinematic viscosity of the He~I layer surrounding the wire, and
$V_\text{eff}=(\rho_n v_n+\rho_s v_s)/\rho$ is the momentum velocity
impinging on the He~I thermal layer, resulting from the interaction
between the external
co-flow at velocity $v_\infty$ and the local counter-flow generated by the heater.

In the King's law, for classical fluids a Prandtl number correction 
$\text{Pr}^{1/3}$ is sometimes
included in the second term of Eq.~(\ref{eq:Nusselt}),
but since this fluid's property is close
to unity for helium in the range of pressures and temperatures of
interest, this correction is not included in our simplified model.

From Eqs.~(\ref{eq:toto1}), (\ref{eq:toto2}), and (\ref{eq:Nusselt}) it follows that 
the heat flux at arbitrary (subsonic) velocity can now be written as
\begin{equation}
  \label{eq:Y1}
  \Phi(T_w , T_\infty, v_\infty) \approx \frac{1}{1+ K \text{Nu}^\star} 
  \left[ \text{Nu}^\star \Phi_I(T_w ) + \Phi_{II}( T_\infty)  \right].
\end{equation}

The velocity dependence is better evidenced by subtracting the heat
flux $\Phi$ in the zero  velocity limit $v_\infty \rightarrow 0^+$.
Retaining the first-order Kapitza correction, we obtain
\begin{equation}
  \label{eq:PhiVOLD}
  \begin{split}
    \Delta\Phi &= \Phi(T_w , T_\infty, v_\infty) - \Phi(T_w , T_\infty, 0^+)   \\
    &\approx B \text{Re}_{w}^{1/2} \left\{  \Phi_I
      -   K \left[  (2A+B \text{Re}_{w}^{1/2}) \Phi_I + \Phi_{II}   \right] \right\}\,.
  \end{split}
\end{equation}

At low enough velocity or temperature (e.g. below $\sim \SI{0.1}{m/s}$ or
below $\sim \SI{2}{K}$, respectively), $\Phi_{II}$ is significantly larger than
$\Phi_I^\star$, and Eq.~(\ref{eq:PhiVOLD}) can be further simplified to obtain
\begin{equation}
  \Delta\Phi  \approx B \text{Re}_{w}^{1/2} \left[  \Phi_I( T_w)   -   K \Phi_{II}( T_\infty)   \right].
  \label{eq:PhiVOLD-simplified}
\end{equation}

Having assumed that the effective velocity $V_\text{eff}$  perceived by
the He~I layer surrounding the wire heater is proportional to
$v_\infty$, we recover the expected $v_\infty^{1/2}$ dependence of
heat transfer rate.

In Eq.~(\ref{eq:PhiVOLD-simplified}), the term within square brackets
increases monotonically with $T_\infty$. Therefore, the behavior with 
temperature of this term alone cannot explain the
observed nonmonotonic dependence of the sensitivity to velocity [see
$\beta(T_\infty)$ in the inset of Fig.~\ref{fig:calibHeII_pow}].
The temperature dependence of the effective
velocity impinging on the wire $V_\text{eff}(v_\infty, T_\infty)$ must
therefore contribute to this dependence, but this remains to be
understood.

\subsection{Local heating in a co-flow and wing bluff bodies}
\label{sec:wings}

This section addresses the flow patterns forming around the heater.
We show that the  flow on a heating wire resembles the flows
on a symmetrical wing:
on its leading edge for the normal-fluid, and its trailing edge for
the superfluid. Each \textit{virtual wing} is characterized by the
two thickness length scales, respectively $L_n$ and $L_s=L_n \rho_n /
\rho_s$, that are significantly larger that the wire diameter in our
experimental conditions.

As a first step, the flows of the normal and superfluid components around a
wire are modeled as two-dimensional potential flows in the plane
perpendicular to the wire, the latter modeled as an infinitely long
cylinder of radius $r_w$. The velocity potential $\Psi (r,\theta)$ of 
the flow around a cylinder is well known 
(see e.g. Ref.~\cite{Batchelor2000}): 
\begin{equation}
  \label{eq:VeloPot1}
  \Psi (r,\theta)=v_\infty \cdot r \left( 1+\frac{r_w^2}{r^2}\right)\cos\theta,
\end{equation}
where $r$ and $\theta$ are polar coordinates whose origin coincides
with the axis of the cylindrical wire, and where the flow far from 
the origin is uniform along the $x$-direction with velocity  $v_\infty$. 
This velocity potential
accounts for the flows of both the normal and superfluid components of the 
external co-flow. A radial local counter-flow from a heating wire can also be
described by the normal and superfluid velocity potentials $\Psi_n$ and $\Psi_s$:
\begin{equation}
  \label{eq:VeloPot2}
  \begin{split}
   \Psi_n (r,\theta) &= \frac{\Phi}{\rho S T} \ln \frac{r}{r_w},\\
   \Psi_s (r,\theta) &=  -  \frac{\rho_n}{\rho_s } \Psi_n (r,\theta).
\end{split}
\end{equation}

An analytical description of the normal and superfluid velocity
fields ($\vvn$ and $\vvs$, respectively) is then obtained 
by superpositions of the local counter-flow potentials~(\ref{eq:VeloPot2})
with the co-flow potential~(\ref{eq:VeloPot1}): 
$\vvn=\bnabla (\Psi  + \Psi_n )$ and $\vvs=\bnabla (
\Psi  + \Psi_s)$.

\begin{figure}[!ht]
  \centering
  \includegraphics[width=1.0\columnwidth]{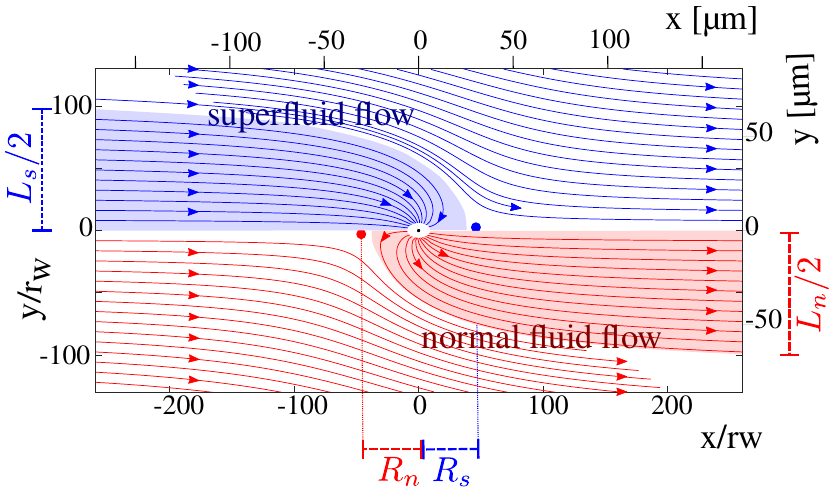}
  \caption{Streamlines of the two-dimensional, normal, and
    superfluid potential flows around a cylinder of radius 
    $r_w=\SI{650}{nm}$ acting as a sink 
    for superfluid (in blue, top half) and a source for the normal fluid
    (in red, bottom half). Each flow is symmetrical with respect to the axis
    $y=0$. The velocity far from the cylinder
    is $v_\infty=\SI{0.25}{m/s}$, and the sink/source properties
    match the local counter-flow produced experimentally for a superfluid
    fraction of 51\%  ($T_\infty=\SI{1.93}{K}$) with $2\pi \Phi=\SI{8.5}{W/m}$,
    which corresponds to a wire overheating around $T_w=\SI{25}{K}$. The
    dimensionless and dimensional scales on both $x$ and $y$ axes are
    relevant for both fluids.}
  \label{fig:puitSource}
\end{figure}

Figure~\ref{fig:puitSource} illustrates the streamlines of the
superfluid (in blue, upper half of the panel) and normal fluid (in red,
lower half) obtained from this model after matching the
cylinder's radius and the mass-flow at the boundaries with the experimental
conditions: a radius $r_w=\SI{650}{nm}$, a co-flow external velocity
$v_\infty=\SI{0.25}{m/s}$, a superfluid fraction of 51\%  ($T_\infty=\SI{1.93}{K}$),
a heating rate per unit length $2\pi \Phi=\SI{8.5}{W/m}$ 
(corresponding to the wire overheating $T_w\approx
\SI{25}{K}$). The background color  highlights the flow regions with
streamlines ending or starting at the surface of the heater.

The length scales of the normal and superfluid flow patterns, $L_n$
and $L_s$, respectively, are about two decades larger than the
wire's radius ($L_n\approx L_s\approx \SI{143}{\mu m}$, see
figure). Henceforth $L_n$ and $L_s$ are called the
heater outer-flow scales.

On the (bottom) normal-fluid side of Fig.~\ref{fig:puitSource}, the
streamlines can be separated into those that are sourced by the
heater, in the red background region, and the others. The first ones
are within the flow ``tail'' of transverse  length scale $L_n$ at $x=\infty$
(see Figure), which can be calculated from the thermal energy balance 
$L_n  v_\infty\rho S T  = 2 \pi \Phi$, where $S$ is the specific entropy 
of the fluid, that is:
\begin{equation}
  \label{eq:Ln}
  L_n    = \frac{2 \pi \Phi}{ \rho S T v_\infty}.
\end{equation}

A normal-fluid stagnation point forms upstream from the wire, at a
distance $R_n$ calculated from the condition $v_n(r,\theta)=0$  for
$r=R_n$ and $\theta=\pi$:
\begin{equation}
\frac{\Phi}{\rho S T R_n}= v_\infty \left( 1- \frac{r_w^2}{R_n^2}\right)
\approx  v_\infty,
\end{equation}
that is:
\begin{equation}
R_n \approx \frac{\Phi}{\rho S T v_\infty } = \frac{L_n}{2 \pi}.
\end{equation}

The  flow of normal fluid outside the red region experiences a
deflection similar to the one on the leading edge of a free-slip
symmetrical wing of thickness $L_n$.

The blue background region on the (top) superfluid side of
Fig.~\ref{fig:puitSource} shows streamlines ``absorbed'' by the heater
surface. The superfluid flow outside this region
experiences a sort of smooth backward step that resembles the flow
in the vicinity of
the trailing edge of a free-slip symmetrical wing. 
The thickness $L_s$ of this ``superfluid wing'' can be calculated from
the mass  conservation to yield
\begin{equation}
 \label{eq:Ls}
 L_s   =  \frac{\rho_n }{\rho_s } L_n, 
\end{equation}
and the position of the superfluid stagnation point, $R_s$, can be
obtained by analogy with the case of the normal fluid as
\begin{equation}
  R_s \approx \frac{\rho_n }{\rho_s } R_n
  = \frac{\rho_n }{\rho_s } \frac{L_n}{2 \pi }= \frac{L_s}{2\pi}.
\end{equation}

\begin{figure*}[!ht]
  \centering
  \includegraphics[width=1.0\textwidth]{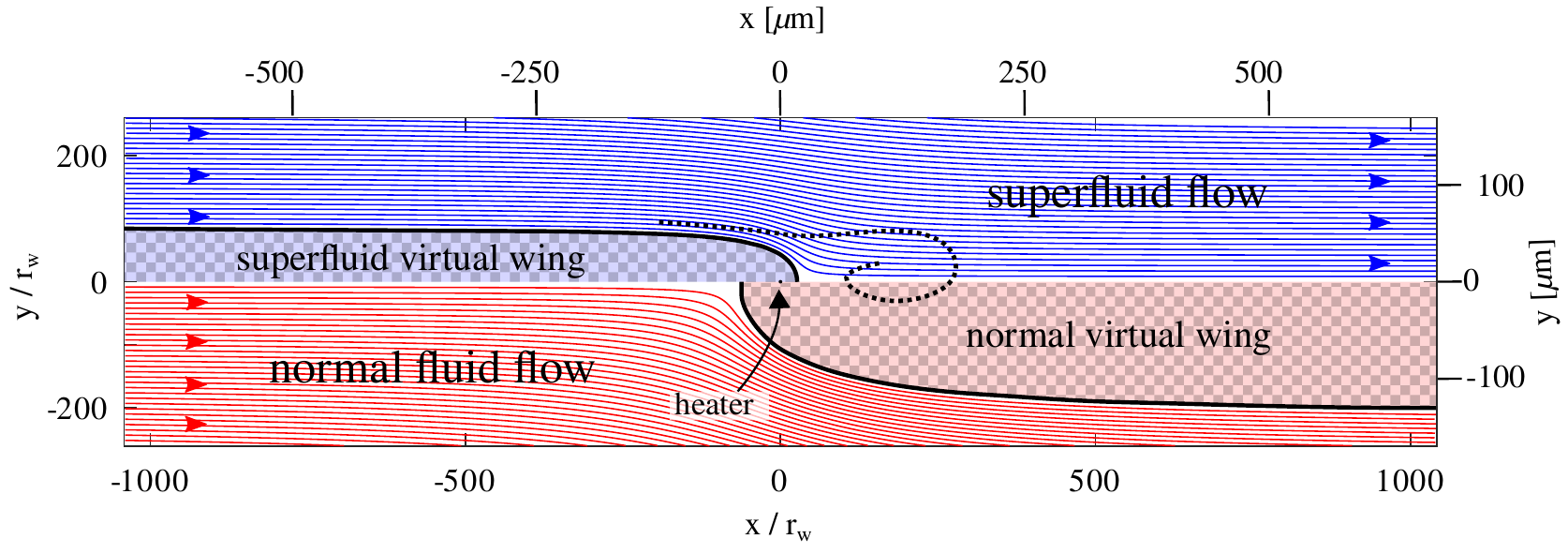}
  \caption{Streamlines of the two-dimensional potential flows around 
    a heater of radius $r_w=\SI{650}{nm}$, for
    $v_\infty=\SI{0.25}{m/s}$, a superfluid fraction of 71\%
    ($T_\infty=\SI{1.74}{K}$) and a heating rate $2\pi
    \Phi=\SI{8.5}{W/m}$ ($T_w\approx \SI{25}{K}$). The region of likely
    recirculation and instabilities is indicated with symbolic
    swirling streamlines. With superfluid-normal fluid coupling (not
    included here), the key hydrodynamic patterns (stagnation zones,
    contours of the virtual wings) are expected to be shifted and
    become time dependent, but we argue that their existence is a
    robust and generic consequence of the local heating in a co-flow.}
  \label{fig:puitSource2}
\end{figure*}

The simple model described in this subsection preserves the key features 
of the normal and superfluid flows in the wide range of conditions
explored. Thus, Fig.\ref{fig:puitSource2} illustrates the flow patterns
around the heating wire of radius $r_w=\SI{650}{nm}$ for 
$T_\infty=\SI{1.74}{K}$. Note that in the considered example the 
length scales of hydrodynamics patterns, whose dependence on physical
parameters is given by Eqs.~(\ref{eq:Ln}) and (\ref{eq:Ls}), remain 
significantly larger than the heater radius.

\subsection{Beyond the model of potential flows}

We  address now the limits of validity of the potential flow
model developed in the previous subsection and analyze the effects of
compressibility, viscosity, mutual friction, and vorticity that have been
ignored so far. We show that the model developed above in
Sec.~\ref{sec:wings} leads, nevertheless, to robust predictions
for the outer-flow patterns at distances from the
heater of the order of or larger than $R_n$ and $R_s$.
Henceforth the flow in the vicinity of the heater  will be called the 
``near-wire flow''.

First it is important to stress that a heater in He~II acts as a sink
of the superfluid component mass flow and source of normal component mass
flow, regardless of the potential-flow modeling.  Thus the existence
of superfluid flow pattern, of typical thickness $L_s$ [see
Eq.~(\ref{eq:Ls})], resembling the trailing-edge of a wing 
is expected to be a
robust feature of the flow, irrespective of modeling. The existence of
a wing-leading-edge pattern of typical thickness $L_n$ [given by
Eq.~(\ref{eq:Ln})] is also a robust feature but we will argue in the
next subsection that its downstream shape probably resembles 
more a wiggling tail than that represented by nearly straight
streamlines. For a point heater, the concept
can be generalized straightforwardly with virtual obstacles having the
shapes of three-dimensional fuselages rather than 
two-dimensional wings.

\textit{Incompressibility}. By definition of length scales $R_s$ and
$R_n$, the velocities at such typical distances from the wire and
beyond are of order $v_\infty$, which, in the conditions typical of the
experiment described above in this paper, is always significantly smaller
than the lowest values of the first and the second sound velocities 
in He~II (respectively $\SI{244}{m/s}$ and
$\SI{6.5}{m/s}$ at $\SI{3}{bar}$ and $\SI{2.13}{K}$). Describing the
flows by incompressible potential fields is therefore justified for 
the outer flow and partly for the near-wire region. 

\textit{Viscosity}. Potential flows are irrotational and thus 
the model developed in Sec.~\ref{sec:wings} is not expected 
to be valid in the wire boundary layer due
to viscous friction of the normal fluid. Nevertheless, compared to
inertial effects, viscous effects are no longer prevalent in the outer 
flow field far enough from the wire. For instance, the relative weakness
of viscous effects at distance $L_n/2$ for the wire can be assessed
from normal fluid Reynolds number $\text{Re}_{L_n}=L_n v_\infty \rho_n /
\mu$, where $\mu$ is the dynamic viscosity of He~I. This Reynolds 
number reaches its smallest values at larger temperature,
where it indeed satisfies the requirement $\text{Re}_{L_n} \gg 1$ 
(e.g., we find $\text{Re}_{L_n} \gtrsim 492$ for $T_\infty=\SI{2.13}{K}$ 
and $2\pi \Phi\gtrsim \SI{2}{W/m}$). Thus, the flow patterns can be estimated
neglecting the normal fluid viscosity in the outer flow and partly in
the near-wire region. 

\textit{Mutual coupling and vorticity}. The chosen velocity potentials
$ (\Psi  + \Psi_n )$ and $ ( \Psi  + \Psi_s)$ describe uncoupled
superfluid and normal fluid. In reality, the presence of superfluid
vortices in the flow is responsible for the mutual friction between
the two fluids,
and eventually a strong coupling of their velocity fluctuations at
scales significantly larger than the typical distance between
superfluid vortices~\cite{Vinen:JLTP2002}. Below we will discuss
in turn the following three flow regions: the upstream region, the
close vicinity of the wire, and the region downstream of the flow.

Upstream, the inter-vortex distance $\delta_\text{co-flow}$ in the
weakly turbulent grid co-flow can be estimated from the
turbulence intensity ($I \approx 2.6\%$, see 
Sec.~\ref{sec:flows}), the turbulent integral
length (say $ L_f \approx \SI{5}{mm}$), and the effective viscosity
$\nu_\text{eff}$ as
\begin{equation}
  \label{eq:intervortex}
  \delta_\text{co-flow}
  \approx  \left( \frac{ \nu_\text{eff} \kappa^2 L_f}
    { I^3 v_\infty^3}\right)^{1/4}
  \approx \SI{40}{\mu m}.
\end{equation}

This formula has been validated by a number of studies
\cite{Salort:EPL2011,Babuin:EPL2014,roche:ETC15}. 
The effective viscosity $\nu_\text{eff}$ is an empirical
quantity defined by postulating that
$\epsilon = \nu_\text{eff}\kappa^2 \mathcal{L}^2$, where $\epsilon$ is
the turbulence dissipation rate and $\mathcal{L}$ the average
superfluid vortex line density~\cite{Niemela05}.
In the considered range of temperatures, $\nu_\text{eff}$ can be 
estimated from experimental values at saturated vapor pressure (see, 
e.g., Refs.~\cite{Niemela05,Babuin:EPL2014}) as $\nu_\text{eff} 
\approx 10^{-8} - 10^{-7} \si{m^2/s}$, or just assuming for 
$\nu_\text{eff}$ the value $\nu_\text{eff} \approx
\mu / \rho  \approx  \SI{1.2e-8}{m^2/s}$ valid for the kinematic
viscosity of the laminar He~II flow, with the dynamic viscosity $\mu$
tabulated in Ref.~\cite{DonnellyBarenghi1998}, or, based on the model
of Ref.~[\onlinecite{Roche2fluidCascade:EPL2009}], as $\nu_\text{eff} \approx
\rho_n B \kappa / (2 \rho)$, where $B$ is a 
tabulated mutual friction coefficient of
order unity~\cite{DonnellyBarenghi1998}. Although those values can
differ by one decade, they all lead to rather close estimates 
for $\delta_\text{co-flow}$ due to the
${1}/{4}$ power law dependence in Eq.~(\ref{eq:intervortex}).

The order of magnitude of $\delta_\text{co-flow}$ is comparable 
to the characteristic scales of the outer-flow, which implies that the
superfluid and the normal fluid are nearly  uncoupled at such scales
before entering the counter-flow  region. Besides, the residual
vorticity associated with the flow's turbulent background hardly distorts
the streamlines due to the weak turbulence intensity. In this regard,
the potential-flow picture is substantiated upstream from the heater. 

In the wire's vicinity, the counter-flow velocities exceed the co-flow
velocity $v_\infty$ and produce a dense turbulent tangle of superfluid
vortices.
The typical intervortex distances $\delta_\text{ctr-flow}$ within
this tangle can be estimated from a well-known thermal counter-flow
equation. Omitting an offset velocity, only relevant at low velocities,
this equation can be written in the form
\begin{equation}
  \label{eq:intervortexCounter}
\delta_\text{ctr-flow}(r) =  \frac{1}{\sqrt{a}| v_s(r) - v_n(r)|},
\end{equation}
where $\vvs \cdot \vvn<0$ and $a(T)$ is a numerical coefficient
tabulated in the literature (see e.g. Ref.~[\onlinecite{Tough1982}]).
Substituting the counter-flow velocities $v_n=\partial \Psi_n (r) / \partial r$ and $v_s=-v_n \rho_n / \rho_s$, one obtains
\begin{equation}
  \label{eq:intervortexCounter2}
  \delta_\text{ctr-flow}(r) =\frac{r}{L_n/2}
  \frac{\pi}{v_\infty \sqrt{a} (1+ \rho_n/ \rho_s) }.
\end{equation}

The counter-flow velocities match in strength the external velocity
$v_\infty$ typically at one outer-scale distances from the
wire. Equation~(\ref{eq:intervortexCounter2}) is no longer strictly valid at
such distance but it should still provide an order of magnitude 
estimate for the
typical intervortex distance upstream from the heater or in the
transverse direction. For instance, for the experimental conditions
modeled in Fig.\ref{fig:puitSource} (\SI{1.93}{K}, \SI{3}{bars}, and
$v_\infty=\SI{0.25}{ m/s}$), we find
\begin{equation*}
  \delta_\text{ctr-flow}(L_n/2)
  \approx \delta_\text{ctr-flow}(L_s/2)
  \approx \SI{1.4}{\mu m},
\end{equation*}
which is two decades smaller than $L_n \approx L_s \approx \SI{140}{ \mu m}$.

We now question if this tangle is dense enough to enforce a significant
coupling between the superfluid and the normal fluid at the outer-flow 
scales $L_s$ and $L_n$. Owing to the large scale separation between
$\delta_\text{ctr-flow}$ and $L_s,L_n$, the superfluid can be described
as a continuous medium characterized by a local vortex line density
$\delta_\text{ctr-flow}^{-2}$ and a coarse grained velocity
$\widetilde{\vvs}$. In such conditions, the coupling between the superfluid
and the normal components can be described, as first approximation,
by a volumetric mutual friction force whose magnitude, $F_{ns}$ can be
written in the Görter-Mellink form
$$F_{ns}=\frac{\rho_n \rho_s}{\rho} \frac{B}{2}\kappa \delta_\text{ctr-flow}^{-2} | \vvn - \widetilde{\vvs} |.$$

Making use of the coarse-grained Hall-Vinen-Bekarevich-Khalatnikov
equations (see e.g. Ref.~\cite{Khalatnikov65}), it is straightforward
to identify the relaxation times  $\tau_n$ and $\tau_s$, due to the mutual friction force, for the normal and superfluid components, respectively, from the estimates for the material derivatives $\vert\rho_nD\vvn/Dt\vert\sim F_{ns}$ and $\vert\rho_sD\vvs/Dt\vert\sim F_{ns}$:
\begin{equation}
  \label{eq:relax}
 \tau_n(r) = \frac{\rho_n }{ \rho_s}  \tau_s(r) = \frac{2 \rho}{B \kappa \rho_s}\delta_\text{ctr-flow}^2(r).
\end{equation}

The superfluid coarse grained velocity $\widetilde{\vvs}$ at a distance
$\sim L_s/2$ from the heater evolves with the characteristic time scale
$L_s/(2v_\infty)$. The mutual coupling will alter significantly the
normal fluid streamlines if the relaxation time $\tau_n(L_s/2)$ is short
enough, say $\tau_n(L_s/2) \lesssim L_s/(2v_\infty)$. Similarly,
mutual coupling will alter the (coarse-grained) superfluid streamlines
at a distance of order $L_n/2$ if $\tau_s(L_n/2) \lesssim  L_n/2v_\infty$.
In the experimental conditions modeled in Fig.~\ref{fig:puitSource}
(1.93 K, 3 bars, $v_\infty=\SI{0.25}{m/s}$, $B\approx1$,
$\rho  \approx 2\rho_s \approx  2\rho_n $), both inequalities become
identical and are found to be valid:

$$ \SI{78}{\mu s} \approx \frac{4}{B \kappa} \left[ \delta_\text{ctr-flow}\left(\frac{L_s}{2}\right) \right] ^2 \lesssim \frac{L_s  }{2 v_\infty}
 \approx \SI{280}{\mu s}. $$

More generally, using Eqs.~(\ref{eq:Ln}), (\ref{eq:Ls}),
(\ref{eq:intervortexCounter2}), and (\ref{eq:relax}), the criteria for
partial fluid locking reduce to

\begin{equation}
  \label{eq:locking}
 \frac{2 \pi \rho S T}{B \kappa a(T)} \frac{\rho_n}{\rho } \left[\max\left(1 , \,
 {\frac{\rho_s}{\rho_n}}\right)\right]^2  \lesssim \Phi.
\end{equation}

Interestingly, this  locking condition amounts to comparing the heat flux with a
quantity that depends only on the helium properties. To the best of our knowledge,
the empirical, temperature-dependent coefficient $a(T)$ is not tabulated in
pressurized helium but the full left-hand-side term can be estimated
at saturated vapor pressure, and it is found to have roughly the same
magnitude as the right-hand-side term of condition~(\ref{eq:locking}),
$\Phi$ shown in Fig.~\ref{fig:null_vel_sato}. 

This shows that the dense superfluid vortex tangle around the heater must
strongly couple the superfluid and the normal components over length scales
encompassing  the outer-flow scales $L_s$ and $L_n$, an effect  ignored in
our simple velocity-potentials model. We thus expect some distortion of the
streamlines, shown in Figs.~\ref{fig:puitSource} and \ref{fig:puitSource2}
within a few $L_s$ and $L_n$ from the heater. Besides, the strong mutual
coupling will favor a locking of the wakes of both fluids and allow vortical
structures to develop in the wake of the heater, definitely invalidating the
model of the irrotational, potential velocity fields downstream from the heater.
The issue of the turbulent wake that forms downstream of the heater is addressed
in the next subsection.

\subsection{The turbulent wake of the heater}

In the previous subsection, we predicted two consequences of a
localized heating in a quantum flow. First, the emergence of virtual
obstacles of typical size $L_s \gg r_w$ (for the superfluid) and $L_n
\gg r_w$ (for the normal fluid) across the flow. Second, a strong
coupling of the superfluid and the normal component flows at length scales
of the order and exceeding $\max(L_s,\,L_n)$; however, in
the vicinity of the heater [say for $r\lesssim
\max(L_s/2,L_n/2)$] the counter-flow velocities remain
significant so that the two fluids tend to move in opposite directions. 

Numerical simulations are probably needed to explore the resulting
hydrodynamic patterns but this is beyond the scope of this
study. Nevertheless, based on a few simple hypotheses we can  
assess the flow stability. First we assume that the outer-flow stability is
controlled by a wake Reynolds number $\text{Re}_\text{ctr-flow}$.
As the superfluid ``wing trailing-edge'' profile is possibly destabilized
at distances of the order $L_s$ from the heater (symbolized 
by the curvy streamline of Fig.~\ref{fig:puitSource2}), we now define the Reynolds number
$\text{Re}_\text{ctr-flow}$ of the flow based on the characteristic length $L_s$. 
As we lack a better understanding of the interplay between the superfluid and the
normal fluid wakes, such a choice of the  length scale to satisfy the conditions of
stability is rather conservative ($\max(L_s,\,L_n)$ would be a less
conservative choice; however, such a choice would not change our quantitative
conclusions). 
At distances of the order $\max(L_n,\,L_s)$ in the wake of the wire,
the counter-flow velocity is small compared to $v_\infty$ but the
vortex tangle still remains dense, thus entailing some re-locking of
the superfluid and the normal fluid velocity fluctuations at scales larger
than the intervortex distance. He~II can then be described as a single
fluid of density $\rho=\rho_s+\rho_n$ and velocity $v \approx v_s
\approx v_n$ that inherits the viscous volumetric force of the normal
fluid $\mu \nabla^2 \vvn \approx \mu \nabla^2 {\bf v}$. The kinematic
viscosity $\nu=\mu/\rho$ of this fluid is thus a natural choice for
the denominator of $\text{Re}_\text{ctr-flow}$. Hence, the wake Reynolds number
defined to assess the stability of the outer-flow  is

\begin{equation}
  \label{eq:Recf}
  \text{Re}_\text{ctr-flow}=\frac{L_s v_\infty }{ \mu / \rho}
  = \frac{\rho_n}{\rho_s} \frac{2\pi \Phi}{ S T \mu}\,.
\end{equation}

Interestingly, this Reynolds number depends only weakly, through 
$\Phi(v_\infty)$, on the velocity of the
external co-flow, $v_\infty$. Formally, Eq.~(\ref{eq:Recf}) reduces to a Reynolds
number characterizing the counter-flow generated by the heater. It can
be formally written using the wire diameter $2r_w$ as the
characteristic length and a characteristic velocity proportional to
the counter-flow superfluid velocity $v_s=\partial \Psi_s (r)
/ \partial r$ extrapolated at the surface of the wire ($r=r_w$):

\begin{equation}
  \label{eq:Recf2}
\text{Re}_\text{ctr-flow}=\frac{(2r_w) [\pi  v_s(r_w)] }{ \mu / \rho}\,.
\end{equation}

In present experimental conditions, $\text{Re}_\text{ctr-flow}$  reaches a few
thousands, far beyond the instability threshold of a flow behind
standard bluff bodies, which becomes unsteady typically for Reynolds
number of a few tens. To summarize, the heater is expected to
generate a turbulent wake of the locked superfluid and the normal fluid and
having a characteristic Reynolds number weakly dependent on the
co-flow velocity $v_\infty$.

\subsection{The vortex street}

In classical hydrodynamics, periodic large scale eddies can form in
the wake of a bluff body~\cite{benard1908vortex}. These structures,
sometimes called von Kármán vortex streets, are characterized by their
shedding frequency $f=\text{St}\cdot U / D$ where $U$ is the flow 
velocity, $D$
is a characteristic transverse size of the obstacle, and 
$\text{St}$ is the
Strouhal number~\cite{strouhal1878} of the order 0.1 -- 0.3 determined 
by the obstacle shape and the Reynolds number based on $U$ and $D$. 
In a turbulent flow, the periodicity of vortex shedding can be altered
and its frequency is not well defined (see,
e.g., Ref.~[\onlinecite{parkerWelsh1983}]). 

For a cylindrical obstacle of diameter $D$, 
$\text{St}\approx 0.20 \pm 0.03$
over the range of Reynolds number 
$DU/\nu \approx 10^2 -2\times10^5$. For a
symmetrical flat wing of thickness $D$ with semicircular leading and
trailing edges, the Strouhal number is slightly larger, e.g., 
$\text{St}\approx 0.27$ for the
aspect ratio 10 and the Reynolds number of 1300~\cite{Nguyen1991}. The
latter geometry is not directly comparable to ours due to the absence
of boundary layer along the superfluid virtual wing. 

Since this vortex shedding effect is inertial and not
viscous~\cite{karmanVortexStreet1912,blevins1984review}, it is
expected to exist in superfluids, although not observed yet to the
best of our knowledge. Could a vortex street account for the spectral
bump at frequency $f_p$ of the heat transfer measurement reported 
above in Sec.~\ref{sec:fglitch}? In other words, could 
cooling glitches be triggered by the shedding of vortices in the wake 
of the heater?

Qualitatively, the absence of the spectral bump at zero velocity is
consistent with this hypothesis. The profile of typical individual
cooling glitches, illustrated in Fig.~\ref{fig:spot_pulses}(b), is also
consistent with emergence of a nonclassical boundary layer, attached 
to the heater (in the form of either the He~I shell, or/and the 
superfluid vortex tangle around it), which undergoes a
re-formation once the velocity perturbations are advected
away. 

More quantitatively, the vortex shedding frequency predicted by
the Strouhal formula can be compared with the measured frequency of the
``bump''.
$U=v_\infty$ is a natural choice for the characteristic velocity of the flow. For now, the effective transverse  length scale of the bluff body will be denoted $D(T,v_\infty)$. 
The vortical patterns emitted behind a symmetrical bluff body have
vorticity of alternating signs, and the frequency given by the Strouhal
number corresponds to the frequency of emission of a 
vortex-antivortex pair. Both vortices from one pair can trigger a 
glitch so that their characteristic frequency $f_p$ would then be 
twice the Strouhal frequency, $f_p= 2 \times \text{St} U / D$.
This leads to the
following prediction for the frequency of the spectral bump: 
\begin{equation}
  \label{eq:Strouhal0}
f_p = \frac{2 \text{St} v_\infty}{D(T,v_\infty)}\,.
\end{equation}

For numerical estimates, we arbitrarily take an intermediate Strouhal
number between the two values cited above: $\text{St}=0.23$.

Measurements reported in Fig.~\ref{fig:fpeak}(b) are consistent with
a linear velocity dependence of $f_p(v_\infty)$, suggesting a weak
velocity dependence of $D$.
The spectra of Fig. \ref{fig:spectraTvar} are consistent with a
spectral bump frequency increasing by at most few tens of
percents from $\SI{1.74}{K}$ to $\SI{2.05}{K}$, suggesting also a weak
temperature dependence of $D$ over this range.
This leads to a preliminary estimate within $\SI{1.74}{K}-\SI{2.05}{K}$ and $v_\infty<\SI{0.4}{m/s}$:
\begin{equation}
D(T,v_\infty)\simeq \SI{75}{\mu m}\,.
\end{equation}

This length scale is significantly larger than the wire diameter. As shown
in Sec.~\ref{sec:coolingglitchesper2}, we can rule out other geometrical
features of the wire, such as its length, since the spectral peaking above
1 kHz associated with this length scale is also observed with the
geometrically dissimilar film probe.
Moreover, we can reasonably exclude that a fixed length scale smaller than
100 microns is present in the incoming flow, since the smallest
(nearly) velocity-independent flow scale is expected to be the integral scale,
which is about two decades larger.
We show below that a correct order of magnitude for $D$ can be obtained
from the naive hypothesis that the thickness $L_s$ of the destabilizing
trailing edge plays the role of the obstacle transverse  length scale.
Indeed, this leads to
\begin{equation}
  \label{eq:Strouhal}
f_p \sim \frac{2 \text{St} v_\infty}{L_s}=
\frac{\text{St} v_\infty^2}{\pi \Phi} \frac{\rho_s}{\rho_n} \rho S T\,.
\end{equation}

This equation corresponds to the dashed line in
Fig.~\ref{fig:fpeak}(b). Strikingly, this naive estimate
gives the correct order of magnitude for the frequency.

The nearly quadratic $f_p\sim v_\infty^2$ scaling
[neglecting the $\Phi(v_\infty)$ dependence] contrasts with
the apparent scalings $f_p\sim v_\infty$ 
or $f_p\sim v_\infty^{1.4}$ of experimental data
[see Fig.~\ref{fig:fpeak}(b) and its inset], indicating that
this simple model does not fully account for the phenomenology at play.
Two corrections could possibly reduce the disagreement. First,
the hypothesis of a 2D flow is not accurate since the wire length
(450 microns) is not much larger than $L_s$. 
The scaling of $L_s(v_\infty)$ could therefore slightly tends toward
the scaling of the axi-symmetric thermal flows expected for point heaters,
that is $L_s \sim v_\infty^{-1/2}$, which could contribute to explain the
sub-quadratic dependence of $f_p(v_\infty)$. Second, a measurement
bias may also contribute to this apparent disagreement:
an underestimate of the true experimental
bump frequency at the largest velocities due to some spectral cut-off.

Over the range $\SI{1.74}{K}-\SI{2.05}{K}$, Eq.~(\ref{eq:Strouhal})
predicts a 38\% decrease of $f_p$, not compatible with measurements.
This suggests that a model for the effective length scale $D$ depending
both on $L_s$ and $L_n$ would be more relevant~\footnote{Empirically,
the length scale $D=(L_s^{-1} + L_n^{-1})^{-1}$ or  $\min (L_s , L_n)$
would fit better the observed  $f_p(T)$ dependence.}.

Further experimental studies in the range $\SI{2.05}{K}-\SI{2.13}{K}$
and numerical simulations would be interesting to complete and underpin
this vortex shedding model.

\subsection{The peak as a filtered high frequency noise}
We present here an alternative explanation of the observed spectral
bump at high frequency.

\begin{figure}[!ht]
  \centering
  \includegraphics[]{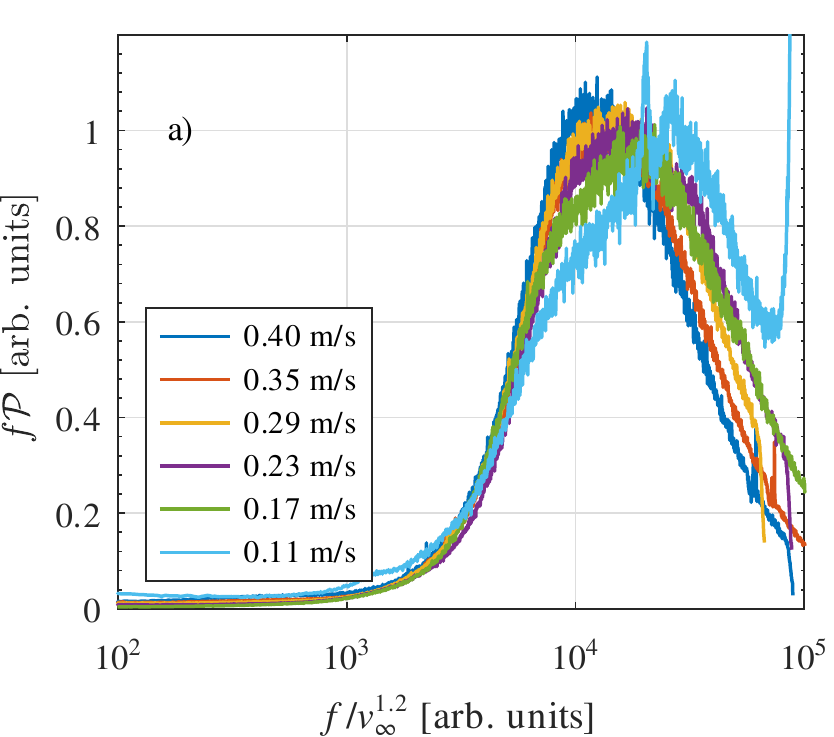}
  \includegraphics[]{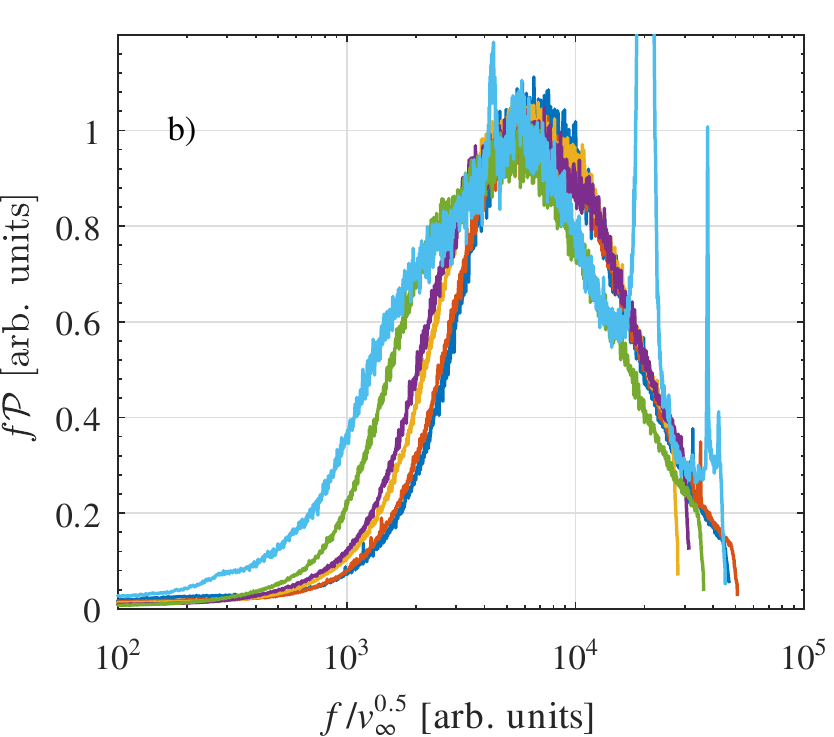}
  \caption{Rescaled power spectral density $f \mathcal{P}(f)$ of the
    wire signal at $\SI{1.74}{K}$
    for various external velocities as a function of the frequency compensated by $v_\infty^\alpha$ with $\alpha = 1.2$ (a) and
    $\alpha = 0.5$ (b). To better illustrate on the left
    and right single curves [panels~(a) and (b), respectively]
    the amplitude of the rescaled PSD was
    normalized for all maxima.}
  \label{fig:enr_peak}
\end{figure}

Instead of characterizing the bump by the frequency of its maximum, as
could have been justified by the resonant or instability mechanisms, 
the bump can
{\it a-priori} be seen as the result of a competition between 
two opposite processes: 
a forcing emerging above some frequency $f_\text{left} < f_p$
(left side of the bump) and a cut-off mechanism at higher frequencies 
(right side of the bump). For instance, the left-side of the bump could be linked to
the onset of processes triggering the cooling glitches, and the
right side to a cut-off associated with a recovery mechanism bounding
the lowest time interval between glitches. It could also be associated
with a finite response time of the heater material and/or the thermal
boundary layer.

Figure~\ref{fig:enr_peak} present spectra in semilogarithmic axis,
with frequency on the $x$ axis compensated by some power of the mean
velocity. On the $y$ axis, the power spectral density $\mathcal{P}(f)$
is multiplied by $f$ so that the surface under the ``curve'' is proportional to
energy $E$ despite the use of logarithmic scale; indeed, for any frequency
interval $\Delta f$ we have

\begin{equation}
  \label{eq:logrep}
  E(\Delta f) = \int_{\Delta  f} \mathcal{P}df
  \propto \int_{\Delta  f}f\mathcal{P} d(\log_{10} f)
\end{equation}

In Fig.~\ref{fig:enr_peak}, the best collapse of the left
(Fig.~\ref{fig:enr_peak}(a)) and right (Fig.~\ref{fig:enr_peak}(b))
hand sides of the bump on single 
curves is obtained by compensating the $x$ axis frequency by $v_\infty^\alpha$ 
with, respectively, $\alpha = 1.2\pm 0.2$ and $\alpha = 0.5\pm 0.05$.

As discussed previously, in the outer flow region the two components
of He~II are expected to become decoupled at scales proportional to the
inter-vortex spacing,  $\delta\propto v_\infty^{-3/4}$. Below this
decoupling scale, it has been predicted that the kinetic energy of the superfluid
component, cascading from larger length scales, will pile up, a phenomenon sometimes
referred to as ``bottlenecking'' or a trend to equipartition~\cite{Salort12}.
At mesoscales this manifests itself as an enhancement of the superfluid velocity fluctuations.
If, by some mechanism yet to be determined, the cooling glitches are triggered by those fluctuations,
then we could expect the formation of the left side of the
peak for frequencies that scale like
$f_\text{left} \propto v_\infty/\delta\propto v_\infty^{7/4}$.

As for the right part of the spectrum, the  $v_\infty^{1/2}$ scaling
could be associated with the thermal response time of the boundary
layer: As the velocity increases, the He~I thermal boundary layer
thickness is expected to scale as $v_\infty^{-1/2}$, and so does the
thermal response time. This yields a cut-off frequency
$f_\text{right} \propto v_\infty^{1/2}$.

\section{Summary and concluding remarks}

Making use of miniature heaters, we have explored the 
forced heat transfer in a subsonic flow of superfluid helium 
at velocities up to $\SI{3}{m/s}$. Our experimental 
results yield the following four main properties of the heat
transfer in He~II flows:

\begin{itemize}

  \item In the case of a sufficiently large overheating of the
  heater, some velocity dependence of the heat transfer rate emerges
  when the fluid in contact with --and in the close vicinity of-- the 
  heater looses its superfluidity.

  \item Two metastable heat transfer regimes exist at large superfluid
  fraction and low velocity. As the velocity is increased, the state
  of lower conduction progressively supersedes  the higher
  conduction state hence leading to a depletion of the mean heat transfer. 

  \item Short-lived cooling enhancements, named \textit{cooling
    glitches}, occur quasiperiodically with a velocity-dependent
  characteristic frequency $f_p$. Their signature in the 
  spectral domain is a broad spectral peak.

  \item Heat transfer sensitivity to velocity reaches a maximum for a
  fluid temperature of $2.0\pm \SI {0.1}{K}$. 

\end{itemize}

An analytical model is proposed to describe these observations. At zero
velocity, it accounts quantitatively for the heat transfer, including
its temperature dependence. At finite velocity, the velocity
dependence is also accounted for, but the maximum, observed at
temperatures around 2~K, of sensitivity of the heat transfer rate
to velocity, remains yet unexplained,
as well as the observation of  metastable states 
at low velocity and low temperature.

A semiquantitative analysis of the flow around the wire heater is
proposed, distinguishing the superfluid and normal fluid components of
the quantum fluid. We predict the formation, around the heater,
of flow patterns in the superfluid and the normal
components whose characteristic scales, respectively $L_s$ and $L_n$,
are two decades larger than the
heater diameter in our experimental configuration. The superfluid
(resp. normal fluid) pattern is reminiscent of the flow over the
trailing  (resp. leading) edge of a symmetrical wing. It is argued
that the dense quantum vortex tangle sustained by the heater couples
the superfluid and normal fluid patterns, resulting in the formation
of a strongly turbulent wake with locked superfluid and normal fluid
components.

The characteristic frequency $f_p$ revealed by heat transfer
measurements is quantitatively consistent with the formation of a von
Kármán vortex street in the wake of the heater. Still, a precise
dependence on the velocity and temperature is not fully accounted for by
the model, calling for further investigations on the relation
between the effective transverse length scale of the obstacle and the 
size of the virtual wings $L_s$ and $L_n$.
We thus discuss an alternative explanation for the appearance
of a broad spectral peak (the so-called ``bump'') at some characteristic
frequency $f_p$ in connection with an existing prediction of a piling up
(or ``bottlenecking'') of the superfluid kinetic energy at small 
scales~\cite{Salort:EPL2011}. We argued that such a peak may result from the
competition between instabilities (cooling glitches) triggered at frequencies
above $f_\text{left} < f_p$ and a cut-off mechanism at higher frequencies.

Numerical studies are certainly needed to better understand the mutual
coupling of the superfluid and normal components in the 
region of local counter-flow generated by the heater, and the relation 
between $L_n$, $L_s$, and the velocity $V_\text{eff}$ perceived by
the He~I boundary layer. 
For instance, it would be interesting to see if the resulting
flow is controlled by the largest or smallest of the two scales
$L_s$ and $L_n$, which
could  explain why the maximum sensitivity to velocity is reached near
2~K, that is when $L_s \approx L_n$.

To conclude, although the heat transfer mechanisms at play are not yet
fully understood and deserve further investigations, 
two applications of the present study can already be suggested. 

First, the miniature heater within the quantum flow can be seen as an
obstacle with a tunable  length scale. Indeed, the length scales $L_s$
and $L_n$ depend on the amount of heating and not on the external
flow velocity. Hence, a three- or two-dimensional
network of miniature heaters
positioned across a flow can be seen as a bluff body with a remotely
controllable shape. This opens an interesting perspective  
in the studies of turbulence generated by an active grid or an active wing. 

Second, a successful operation of the hot wire anemometer in superfluid
has been previously reported~\cite{Duri15}. The present study confirms
the analysis and conclusions of the cited work but also allows us
to identify the following limitation of the hot-wire anemometry in a 
quantum flow: The space-time
resolution is spoiled by the formation of the outer flow scales ($L_s$
and $L_n$) and the associated time scales ($L_s / v_\infty$ and $L_n /
v_\infty$). In particular, the broad peak in the spectral response 
cannot be interpreted directly as a feature of the external flow, and
it would be hazardous to identify it with the predicted bottlenecking of the
velocity spectra in superfluid helium at finite
temperature\cite{Salort:EPL2011,Lvov07}.

\section{Acknowledgments}
We thank Bernard Castaing for providing us with a key step
allowing building the analytical expression for the heat transfer at
null velocity. We also thank Rémi Benhafed and Jérome Chartier for
support in designing an running the experiment.
This work is supported by the European Commission --
European High-performance Infrastructures in Turbulence (CE-EuHIT,
project ``MOVEMENT2''), and the French National Research Agency (ANR) 
Grant No.~09-BLANC-0094 (project ``SHREK'') and
Grant No.~18-CE46-0013-03 (project ``QUTE-HPC'').

\appendix*
\section{Characterization of the grid flow}

In the Appendix we shall use local velocity measurements performed by means of the
wire heater in He~I to compute the integral length scale $L_f$
and the turbulent intensity $I$ of the flow.
There is a wealth of evidence that these large scale flow properties
should remain largely unaffected by the transition from classical
turbulence in He~I to the He~II quantum turbulent flow~\cite{Maurer98,Salort10}. With these
primary quantities, we shall then estimate the Reynolds number and the
so-called Kolmogorov dissipative length scale under the assumption of
homogeneous and isotropic turbulence.

\paragraph{Calibration of the wire heater}
In He~I the wire heater behaves as a conventional hot wire anemometer:
As explained in Sec.~\ref{sec:model_vit}, the King's law is then very
well suited to fit the electrical power dissipated in the wire
heater as a function of the velocity. Here we use the raw King's
law for calibration:
\begin{equation}
  \label{eq:king}
  e^2 = C + Dv_\infty^{0.5},
\end{equation}
where the calibration constants $C$ and $D$ are determined using a
polynomial fit of the mean voltage against the mean velocity in
the tunnel. 

\paragraph{Turbulence intensity}

Here we compute the turbulence intensity $I = v'/v_\infty$ where
$v'$ is the standard deviation of the velocity.

The raw voltage records, and thus the velocity data, are affected by an
uncorrelated base noise. Therefore, at low velocity, in which case 
the signal to noise ratio is small, the evaluation of  the standard deviation of the
velocity is not reliable. It is thus reasonable to use, for the
turbulence intensity, the value found at the highest velocities:
$$I = v'/v_\infty\approx 2.6\%.$$

\begin{figure}[!ht]
  \centering
  \includegraphics[]{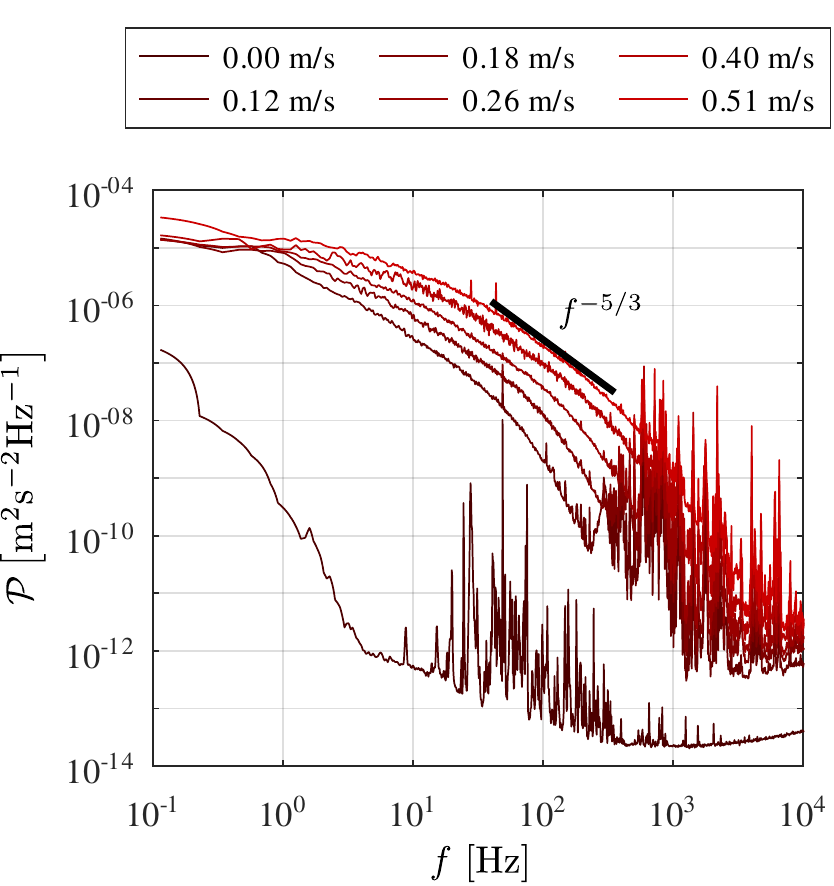}
  \caption{Power spectral density of velocity time
    series as a function of the frequency at flow velocities $v_\infty$ ranging from
    \SI{0}{m.s^{-1}} (dark red) to \SI{0.51}{m.s^{-1}}  (light red).}
    \label{fig:spectra-app}
\end{figure}

\paragraph{Integral length scale}
The longitudinal integral length scale is defined as follows:
\begin{equation}
  \label{eq:Lf}
  L_f = \int_0^\infty \mathcal{R}_{xx}d\delta_x,
\end{equation}
where $\mathcal{R}_{xx}$ is the autocorrelation coefficient of the
longitudinal component of the fluctuating velocity $v'$
along the longitudinal axis $x$
\begin{equation}
  \mathcal{R}_{xx} (\delta_x)
  = \langle v^\prime(x) v^\prime(x+\delta_x)\rangle/
  \langle v^{\prime 2}\rangle.
\end{equation}

Since we have access to the time series of the velocity, we use the Taylor
hypothesis of frozen turbulence in order to transform temporal to
spacial data through $x = \langle v \rangle t$.

The integral~(\ref{eq:Lf}) converges to
$$L_f = 5.0\pm \SI{0.2}{mm}$$
after a length scale $\delta_x$ that ranges between 2000$L_f$ and 10000$L_f$.

\paragraph{Reynolds number and Kolmogorov dissipative length scale.}
As the hot wire only gives access to one component of the velocity 
(streamwise), in order to compute the Reynolds number we first need to
make an assumption on the isotropy and homogeneity of the flow. Then
we can rely on the relation $R_\lambda = \sqrt{15 \text{Re}_{Lf}}$ 
where $\text{Re}_{L_f} = I\langle v\rangle L_f/\nu$ (see
e.g. Ref.~\cite{Batchelor53}). For the  largest velocities, this yields
$$R_\lambda\approx 230.$$

We can also evaluate the Kolmogorov dissipative length scale $l_\eta$
based on the definition $l_\eta/L_f = \text{Re}_{L_f}^{-3/4}$ which leads to
$$l_\eta \approx\SI{10}{\mu m}.$$

\paragraph{General comments on spectral data in He~I}

In order to make general comments on the quality of the acquired fluctuating
velocity time series, we can look at their power spectral density
[PSD or $\mathcal{P}(f)$ hereafter] which has a very well
known spectral signature in the considered grid flow.

As can be seen from Fig.~\ref{fig:spectra-app}, at all available  velocities 
$v_\infty$ the PSD scales with the frequency reasonably well as $f^{-5/3}$. 
The extent of the inertial range in the frequency domain is about one decade, 
which is quite good considering that
the length of the wire is only 10 times smaller than the large scale
of the flow $L_f$. Finally we note that for nonzero velocities there
appears a noise at frequencies  above 200~Hz. We did not manage to
determine the source of this noise, but as it only introduces
a small amount of energy at high frequency, we do not expect it to affect
the conclusions of this paper.

\bibliographystyle{apsrev4-2}
\bibliography{multistable_hotwire}
\end{document}